\documentclass[acmtog,screen,nonacm,balance=false]{acmart}

\usepackage{xcolor}
\usepackage{wrapfig}
\usepackage[noend]{algpseudocode}
\def\BState{\State\hskip-\ALG@thistlm}
\makeatother

\newcommand{\algorithmicforeach}{\textbf{foreach}}

\algdef{SE}[FOREACH]{ForEach}{EndForEach}[1]{\algorithmicforeach\ #1\ \algorithmicdo}{\algorithmicend\ \algorithmicforeach}%
\algtext*{EndForEach}

\AtBeginDocument{}

\usepackage{rotating}

\newcommand{\xx}{\mathbf{x}}
\newcommand{\uu}{\mathbf{u}}
\newcommand{\ww}{\mathbf{w}}

\newcommand{\vvv}{\mathbf{v}}

\newcommand{\qq}{\mathbf{q}}

\newcommand{\cc}{\mathbf{c}}

\newcommand{\BB}{\mathbf{B}}

\newcommand{\ee}{\mathbf{e}}

\newcommand{\CC}{\mathbf{C}}
\newcommand{\RR}{\mathbf{R}}
\newcommand{\rr}{\mathbf{r}}

\newcommand{\OO}{\mathbf{O}}

\newcommand{\dd}{\mathbf{d}}
\newcommand{\pp}{\mathbf{p}}

\renewcommand{\SS}{\mathbf{S}}

\renewcommand{\ss}{\mathbf{s}}

\renewcommand{\gg}{\mathbf{g}}
\newcommand{\GG}{\mathbf{G}}

\renewcommand{\tt}{\mathbf{t}}

\newcommand{\XX}{\mathbf{X}}

\newcommand{\one}{\mathbf{1}}

\newcommand{\QQ}{\mathbf{Q}}

\newcommand{\NN}{\mathbf{N}}
\newcommand{\TT}{\mathbf{T}}
\newcommand{\II}{\mathbf{I}}

\newcommand{\nn}{\mathbf{n}}

\newcommand\restr[2]{{\left.\kern-\nulldelimiterspace{}#1\right|_{#2}}}

\usepackage{bm}
\usepackage{tikz}
\usepackage{xcolor}

\definecolor{mygreen}{RGB}{46,160,67}
\definecolor{myyellow}{RGB}{227,179,65}
\definecolor{cdorange}{RGB}{232,114,28}

\let\oldnl\nl
\newcommand{\nonl}{\renewcommand{\nl}{\let\nl\oldnl}}
\newlength\savedwidth
     
\usepackage[switch]{lineno}
\usepackage[boxed,ruled, linesnumbered]{algorithm2e} 

\SetCommentSty{mycommfont}

\usepackage[ruled]{algorithm2e} 

\SetAlFnt{\small}
\SetAlCapFnt{\small}
\SetAlCapNameFnt{\small}
\SetAlCapHSkip{0pt}

\usepackage[normalem]{ulem}

\DeclareRobustCommand{\del}[1]{}

\DeclareRobustCommand{\rtodo}[2]{}

\newcommand{\lastpagecolumnbreak}{\vadjust{\vfil\penalty-10000}}

\usepackage{tikz}
\usetikzlibrary{angles, calc, decorations.pathreplacing}
\tikzset{
  dot/.style={
    circle, fill=black, inner sep=1pt, outer sep=0pt
  },
  dot label/.style={
    circle, inner sep=0pt, outer sep=1pt
  },
  pics/right angle/.append style={
    /tikz/draw, /tikz/angle radius=5pt
  }
}
\usepackage{xcolor}
\newcommand{\tikzcircle}[2][red,fill=red]{\tikz[baseline=-0.5ex]\draw[#1,radius=#2] (0,0) circle ;}%

\begin{document}
\title{HairCS: Reconstructing Strand-Based Hair from Hair Cards}

\setcounter{footnote}{1}  
\author{Zixuan Lu}
\authornote{This work was done when Zixuan Lu was an intern at LIGHTSPEED.}
\email{birdpeople1984@gmail.com}
\affiliation{%
  \institution{University of Utah}
  \country{United States of America}}

\author{Tongtong Wang}
\email{wangtong923@gmail.com}
\affiliation{%
  \institution{LIGHTSPEED}
  \country{Australia}}

\author{Yuefan Shen}
\email{jhonve@zju.edu.cn}
\affiliation{%
  \institution{LIGHTSPEED}
  \country{China}}

\author{Zhongtian Zheng}
\email{zhengzhongtian@pku.edu.cn}
\affiliation{%
  \institution{LIGHTSPEED}
  \country{China}}

\author{Chenfanfu Jiang}
\email{chenfanfu.Jiang@gmail.com}
\affiliation{%
  \institution{UCLA}
  \country{United States of America}}

\author{Yin Yang}
\email{yangzzzy@gmail.com}
\affiliation{%
  \institution{University of Utah}
  \country{United States of America}}

\author{Kui Wu}
\email{walker.kui.wu@gmail.com}
\affiliation{%
  \institution{LIGHTSPEED}
  \country{United States of America}}

\renewcommand\shortauthors{Lu et al.}
\makeatletter
\let\haircs@authornotes\@authornotes
\def\@authornotes{\setcounter{footnote}{1}\haircs@authornotes}
\makeatother

\begin{abstract}
We present an automated pipeline that converts hair-card models into high-quality strand-based hairstyles. Given a collection of textured triangular or quad strips as input, our method produces a strand-based representation that preserves the original hairstyle while enriching it with fine-scale geometric detail and adhering to standard production requirements: strands originate from the scalp, roots are uniformly distributed, and the hair volume is plausibly filled. The resulting assets are directly compatible with strand-based rendering, physics-based simulation, and common grooming modifiers (e.g., clumping, curling, noise) for enhanced realism and artistic control. We validate our approach on a large and diverse set of hairstyles, including short and long hair, curly styles, and complex styles such as buns and ponytails. 
\end{abstract}

%
%

%
%
\begin{teaserfigure}
\centering
\includegraphics[width=\textwidth]{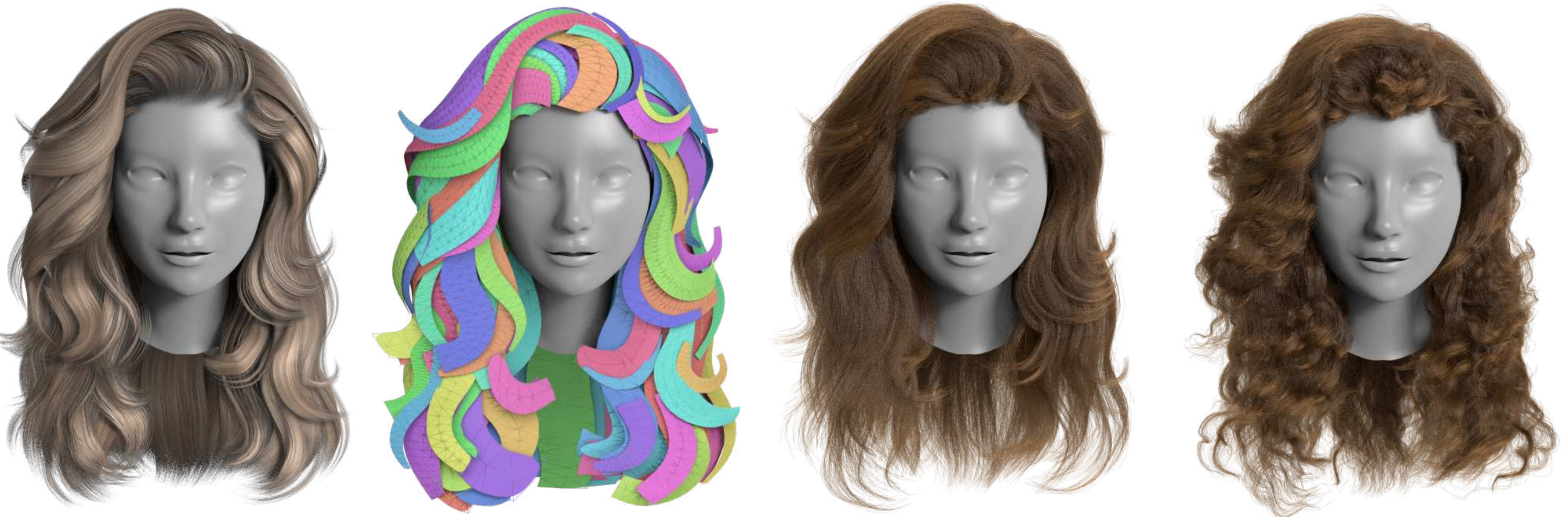}
\captionof{figure}{We present a pipeline that automatically converts a hair-card model into a high-quality strand-based hairstyle. Our five-step pipeline begins with guide extraction and constructs a wrapper geometry that captures the wisp's volumetric occupancy. The resulting strands are compatible with grooming operators and ready for strand-based simulation and rendering in industrial game engines. From left to right: hair cards rendered with their textures, the bare card geometry, our output strand-based model, and our result enriched with the curly operator.}
\label{fig:teaser}
\Description{}
\end{teaserfigure}


\maketitle
\section{Introduction}

Hair modeling and generation are fundamental to the creation of photorealistic digital humans and play a critical role in a wide range of applications, including virtual avatars, AR/VR, games, and digital fashion. However, industrial hair-authoring pipelines remain bottlenecked by labor-intensive manual creation, since hair exhibits rich geometric detail, complex spatial organization, and highly distinctive structures across individuals and hairstyles. As a result, automatic hair generation from single-view and multi-view images has become an active research topic in recent years, with a growing reliance on data-driven and learning-based methods~\cite{zhou2018hairnet,zhang2019hair,wu2022neuralhdhair,zheng2023hairstep,he2025perm,kuang2022deepmvshair,rosu2022neural,sklyarova2023neural,wu2024monohair,takimoto2024dr,zakharov2024human,zhou2024groomcap}.

Although recent methods have achieved substantial progress in reconstruction quality, the scarcity of a large-scale, high-fidelity, and diverse hair dataset continues to limit the performance of data-driven and learning-based approaches. At the same time, strand-based hair simulation and rendering have become increasingly important in modern production pipelines because they capture dynamic behavior and visual richness with high realism~\cite{tafuri2019strand,unrealengine,Hsu2025}. Nevertheless, hair cards, flat textured strips used to approximate the appearance of hair clusters~\cite{Zheng2025card,Tojo2025,Yibing2016}, remain widely adopted in real-time applications due to their efficiency and practicality. Consequently, a large number of high-quality hair assets in existing game and production pipelines are still represented as hair cards.
Motivated by this observation, we aim to convert existing large-scale hair-card assets into a high-fidelity, diverse strand-based hair dataset to support future research in hair reconstruction, generation, and modeling. A second target application of our method is to transform hair-card models originally designed for low-end platforms or mobile games into strand-based hair models with sufficient geometric detail for AAA game production.

As shown in~\autoref{fig:teaser}, this paper presents an automated pipeline that converts input hair-card models into high-quality strand-based hair representations. The input comprises a set of flat, textured strips that approximate the appearance of hair clusters, and the output is a strand-based hair model that preserves the original hairstyle structure while enriching it with fine-scale geometric detail. Our formulation further conforms to the standard strand-based hair setting: each strand grows from the scalp, roots are uniformly distributed over the scalp, and strands spatially fill the hair volume in a plausible manner, thereby enabling strand-based rendering and simulation. In addition, generated hair can be easily enhanced with common grooming modifiers, such as clumping, curling, and fuzz, to improve realism and provide greater artistic control. 

Three difficulties make this conversion more than an interpolation problem: card roots are not distributed over the scalp, flat proxies must be lifted into a hair volume without voids or crowding, and partings and layering must survive. We address them with (i) a parting-aware guide binding, (ii) a wrapper representation relaxed by ACCD expansion with an XPBD rod solver, and (iii) a density-targeted relaxation of the full strand set.

Our pipeline begins by extracting guide hairs from each hair-card strip and redistributing their roots over the scalp by solving an integer programming problem. Next, guide hairs are inflated into wrappers, whose cross-sectional areas are constrained to match their corresponding scalp coverage, via a second optimization procedure designed for efficient relaxation. Each wrapper is then instantiated as a cluster of individual hair strands, thereby forming a complete strand-based hair model. Finally, a third optimization step adjusts the spatial distribution of strands and fills the gaps between neighboring wrappers, yielding a more natural and visually coherent hairstyle. 
We evaluate our method on a large dataset of hairstyles from~\cite{Zheng2024}, covering a wide range of variations, from short to long hair, and from curly styles to complex updos such as buns. Our method can faithfully convert these hair-card models into high-quality strand-based hair representations, enabling several downstream applications, including physics-based strand simulation, the creation of a 50K strand-based hair dataset via hair blending, and secondary editing via grooming modifiers.
The dataset is available at \url{https://huggingface.co/datasets/HairCS2027/HairCS} and is described in \autoref{apdx:dataset}.

\section{Related Work}

\paragraph{Hair Representation}
Hair simulation and rendering are computationally expensive because of the large number of strands and their complex interactions. To reduce this cost, prior work has proposed a range of simplified representations, including 2D strips, hair cards~\cite{Koh2001, ward2003modeling}, cubic lattice structures~\cite{Volino2006hair}, short hair strips~\cite{Guang2002hair}, and volumetric representations~\cite{wu2016hairmesh, Lee2019volumehair}. During rendering, these reduced representations are typically expanded into full hair using baked textures or procedural functions. Among them, hair cards remain widely used in the gaming industry due to their simplicity and efficiency~\cite{Yibing2016}. We refer readers to the course by \citet{Bertails2008} for a comprehensive overview of hair rendering and simulation. More recently, learning-based methods have explored neural representations for reconstructing and representing complex hair geometry~\cite{luo2024gaussianhair, zheng2025groomlight, wang2023neuwigs}.

\paragraph{Hair Modeling}
Hair models are traditionally created by artists using tools such as Maya XGen, but manual authoring remains labor-intensive. To simplify this process, \citet{Yuksel2009} introduced \emph{hair mesh}, a volumetric mesh representation that provides high-level editing controls and generates strands procedurally within the mesh. \citet{Wu2024CurlyCue} presented a geometric method for modeling highly coiled hairstyles such as afros. More recently, \citet{huang2024real} proposed a real-time framework that dynamically groups hair into thick, camera-facing cards for rendering. To reduce the effort required to author hair cards, \citet{Zheng2025card} and \citet{Tojo2025} introduced automatic pipelines that convert strand-based hair models into cards using differentiable rendering. \citet{Chang2025} proposed an inverse hair grooming pipeline that transforms 3D strands into procedural hair grooms that consist of a small set of guide strands and hair grooming operators.

\paragraph{Hair Generation}
Sketch-based interfaces provide intuitive control for hairstyle design by allowing users to draw hair directly in screen space~\cite{fu2007sketching, shen2020deepsketchhair}. For automatic hair generation, many methods reconstruct strand-based geometry from images. Earlier approaches relied on heuristics~\cite{Kong1998generation, paris2008hair, jakob2009capturing, sun2021human, hu2017avatar} or large 3D hairstyle databases~\cite{hu2015single, chai2016autohair, liang2018video}. More recent learning-based methods have improved the fidelity and robustness of reconstruction from single-view~\cite{zhou2018hairnet, zhang2019hair, wu2022neuralhdhair, zheng2023hairstep} and multi-view inputs~\cite{kuang2022deepmvshair, rosu2022neural, sklyarova2023neural, wu2024monohair, takimoto2024dr, zakharov2024human, zhou2024groomcap}. Beyond reconstruction, hair synthesis methods perform hairstyle transfer using feature maps~\cite{Wang09} and, more recently, learned features~\cite{zhou2023groomgen, sklyarova2024text, chen2024doubly, he2025perm}. However, these learning-based methods often rely on slow optimization procedures, or their output quality is limited by inadequate representations, such as low-resolution direction or density grids, and by the restricted hairstyle diversity and quality of the training data.

\paragraph{Hair Dataset}
\citet{hu2015single} introduced USC-HairSalon, the first publicly accessible 3D hairstyle database for data-driven hair modeling research. This dataset has supported hair modeling and generation research for more than a decade and has enabled numerous follow-up studies~\cite{zhou2023groomgen,he2025perm,difflocks2025,wu2024monohair}. More recently, it was extended into Hair20K~\cite{Hair20k}, a large 3D hairstyle database of 20K hairstyles generated using hair-blending techniques. In recent years, researchers have increasingly recognized the importance of high-quality hair datasets, as they strongly influence the performance of learning-based methods. CHARM~\cite{He2025charm}, for example, constructs AnimeHair, a large-scale dataset of 37K high-quality anime hairstyles derived from publicly available VRoid models, where each hairstyle is represented by volumetric pyramids corresponding to individual hair clusters. DiffLocks~\cite{difflocks2025} also released 40K synthetic hairstyles created in Blender, including curly, balding, combed-back, and afro-like styles. CT2Hair~\cite{Shen2023ct2hair} released 10 high-quality strand-based hair models reconstructed from real-world wig CT scan data. UniHair~\cite{Zheng2024} introduced the large-scale  hair dataset SynMvHair, which covers a wide range of hairstyles and contains 2,396 collected 3D hair-card models and 82,682 texture maps in total. These cards are artist-authored game assets collected from The Sims Resource, shipped with shared texture atlases and inconsistent UV conventions. Our method builds on the SynMvHair dataset by converting its hair-card assets into strand-based hair models and further expands it to a 50K strand-based dataset via hair blending  and grooming operators.
\begin{figure*}[ht!]
    \centering
    \includegraphics[width=\linewidth]{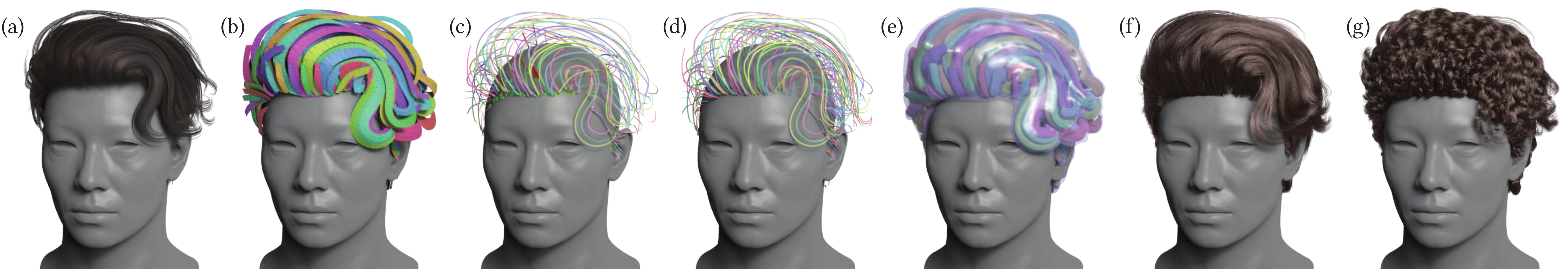}
    \caption{\textbf{Our pipeline.} 
    Given textured triangle strips as hair cards, (a) textured and (b) pure geometry, we extract each card's flow orientation, root side, and centerline, then bind the resulting guides to root candidates on the scalp. (c) shows the binding relaxation, with bound segments highlighted in red. The set of unbound roots is then sparsified, and a group of extra guides is added to cover the entire scalp. The root-connected guides are then smoothed to form the final guide set (d). We construct a wrapper and partition it via an expansion algorithm (e), then synthesize our base hair within the wrapper (f). The resulting base hair can be edited into a variety of styles, such as curly hair, via the groom modifier (g).}
    \label{fig:pipeline}
    \Description{}
    \vspace{-1em}
\end{figure*}

\section{Overview}


\paragraph{Problem Statement.} 
Given a textured hair-card mesh $\mathcal{C} = \{\CC_i\}$ with embedded UV coordinates, a bust mesh $\mathcal{B}$, and a scalp region $\mathcal{S}$ $\subset \mathcal{B}$, our goal is to generate a dense strand-based hair model $\mathcal{S}^{\text{out}} = \{\SS_i\}$. Each strand $\SS_i = \{\ss_{ij}\}$ is represented as a polyline formed by connecting $|\SS_i|$ sample points $\ss_{ij}$, where $\ss_{i0}$ denotes the hair root located on the scalp. The output strand-based model $\mathcal{S}^{\text{out}}$ is expected to preserve the structure of the original hairstyle represented by the hair cards $\mathcal{C}$, while conforming to the standard strand-based hair setting: roots $\{\ss_{i0}\}$ are uniformly distributed over the scalp, and strands $\{\SS_i\}$ plausibly fill the hair volume in space, thereby enabling strand-based rendering and simulation.

The challenges are twofold. First, when authoring hair cards, artists primarily focus on the external visual appearance rather than the underlying strand structure, and the cards are typically not distributed uniformly over the scalp. Second, hair cards are flat surface proxies that approximate clusters of fibers, whereas strand-based hair is represented by volumetric curves. Therefore, the generated strands must not only preserve the original hairstyle, but also uniformly and plausibly fill the hair volume.

\paragraph{Our Pipeline.} \autoref{fig:pipeline} illustrates our overall pipeline. We begin by extracting guide hairs (\emph{guides}) from the input hair cards (\autoref{sec:extraction}), and then bind their roots to uniformly sampled scalp positions (\autoref{sec:binding}). To further obtain a guide set that is more uniformly distributed over the scalp while preserving the structure of the input hair cards, we generate additional guides by tracing from unbound scalp samples selected via farthest-point sampling (FPS) (\autoref{sec:tracing}). The traced guides, together with the guides extracted from the cards, form the complete guide set.
Before synthesizing the final hair strands, each guide is converted into a \emph{wrapper}, which originates from an associated region on the scalp and sweeps along the guide to occupy a volumetric region. To provide an efficient approximation of the spatial hair distribution, we perform wrapper-level relaxation, allowing the wrappers to progressively fill the available hair volume while preserving the initial structure and overall silhouette specified by the input hair cards (\autoref{sec:relaxation}).
Finally, full hair strands are synthesized within each wrapper using parallel transport. The synthesized strands are then further refined by a second, strand-level relaxation to remove visible gaps and discontinuities between hair clusters generated from different wrappers, while remaining inside the hair volume and preserving the initial hairstyle shape (\autoref{sec:full-hair}).

\section{Method}

\subsection{Guide Extraction}\label{sec:extraction}

For each hair card $\CC_i$, which consists of a sequence of quads, we first identify its four boundary corners and treat the card as a rectangle in UV space. Since a single hair texture may encode strand flow along either the $u$-direction or the $v$-direction, we determine its dominant orientation using the 2D orientation-field analysis of~\cite{Paris04}. Specifically, we convert the texture to grayscale and apply a high-pass filter with a wide Gaussian kernel to suppress low-frequency lighting effects, yielding the filtered texture $\II_i$. We then compute a phase-invariant response over the masked texture region, where $\pp \in \II_i$ denotes a pixel belonging to hair, and obtain the feature value $E_i(\theta)$ for each orientation $\theta$:
\begin{equation}\label{eq:gabor_energy}
E_i(\theta) \;=\; \sum_{\pp \in \II_i} \,\sum_{\lambda} \sqrt{\bigl[\bigl(\II_i * G^{\,\mathrm{c}}_{\theta,\lambda}\bigr)(\pp)\bigr]^2 + \bigl[\bigl(\II_i * G^{\,\mathrm{s}}_{\theta,\lambda}\bigr)(\pp)\bigr]^2},
\end{equation}
where $G^{\,\mathrm{c}}_{\theta,\lambda}$ and $G^{\,\mathrm{s}}_{\theta,\lambda}$ are the cosine- and sine-phase Gabor kernels~\cite{Jain91} at orientation $\theta$ and wavelength $\lambda$, respectively, and $*$ denotes the 2D convolution.

\begin{figure}[ht!]
    \centering
    \includegraphics[width=\linewidth]{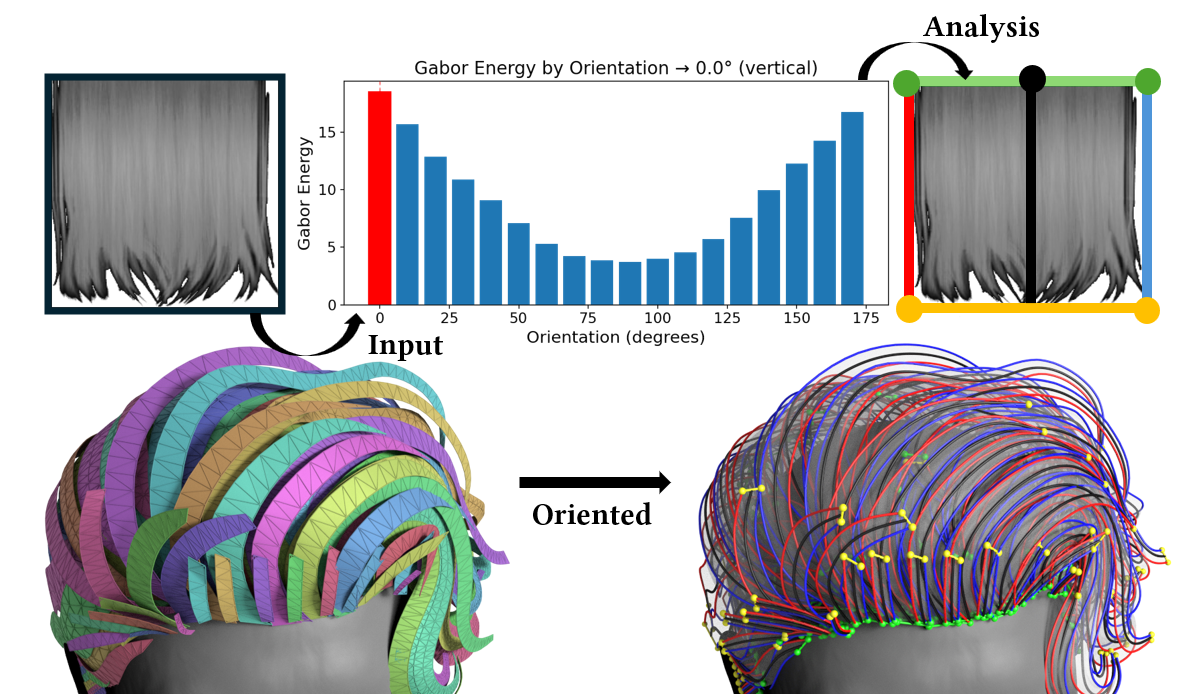}
    \caption{\textbf{Card orientation analysis.} The high-pass filtered grayscale texture of each card is passed through the orientation analysis, which yields the root side (\tikzcircle[mygreen,fill=mygreen]{2.0pt}), the tip side (\tikzcircle[myyellow,fill=myyellow]{2.0pt}), evenly spaced root samples (\tikzcircle[black, fill=black]{2.0pt}), and the flow direction (\tikzcircle[mygreen, fill=mygreen]{2.0pt}$\rightarrow$ \tikzcircle[myyellow, fill=myyellow]{2.0pt}). The oriented 2D information is then lifted into the 3D shape. }
    \Description{}
    \label{fig:extraction}
\end{figure}

The dominant orientation is then given by $\theta^{\star}_i = \arg\max_{\theta} E_i(\theta)$. Because this orientation is symmetric up to sign, we classify it according to its angular distance to the $u$-axis: if $\min(\theta^{\star}_i,\, \pi - \theta^{\star}_i) \le \pi/4$, we infer that the hair flow runs along the $v$-direction; otherwise, it runs along the $u$-direction.  Finally, we generate a set of guides $\mathcal{G} = \{ \GG_i \}$ along the inferred root-to-tip direction at even spacing. The process is shown in~\autoref{fig:extraction}.
The only assumption of this stage is that each card is a UV strip with a monotone flow axis. It assumes no fixed UV orientation, no fixed root side, no single card per texture region, and no flat rectangular strip: the axis is read from each card's texture, and the outlines and guides are lifted to 3D through the card's own UV mapping, so shared atlases and folded strips need no topological analysis. The root is first taken at the start of the strip along the detected axis, and scalp proximity overrides this choice only when the root end is clearly far from the scalp while the tip end is close to it.

\subsection{Guide Binding} \label{sec:binding}

As our ultimate goal is to uniformly distribute hair roots over the scalp, given the extracted guides $\mathcal{G} = \{\GG_i\}$ from \autoref{sec:extraction}, we first distribute guide roots over the scalp region $\mathcal{S}$. Instead of distributing guide roots uniformly over the scalp, which would ruin the hair structure, we pre-sample the scalp $\mathcal{S}$ with $N^{\text{root}} = 30\mathrm{K}$ root candidates $\mathcal{R} = \{\rr_i\}$ using Poisson-disk sampling, and aim to bind the $N^{\text{guide}}$ guides $\mathcal{G}$ to these candidate roots. Note that we over-sample the scalp to ensure that each guide can find the closest roots, but do not require guides to be uniformly distributed.

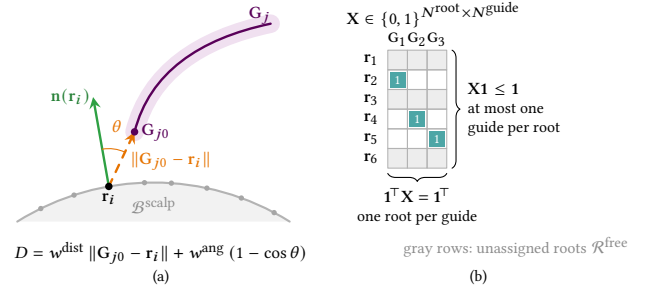
\begin{figure}[ht!]
    \centering
\begin{tikzpicture}[black, every node/.style={font=\scriptsize, inner sep=1.5pt}, >=stealth, line cap=round, line join=round]
  \begin{scope}
    \coordinate (C) at (1.8,-3.0);
    \fill[gray!10] ($(C)+(58:3.4)$) arc[start angle=58, end angle=122, radius=3.4] -- cycle;
    \draw[gray!60, line width=0.8pt] ($(C)+(58:3.4)$) arc[start angle=58, end angle=122, radius=3.4];
    \node[gray!90] at (1.8,0.1) {$\mathcal{B}^{\text{scalp}}$};
    \foreach \a in {62,70,78,86,92,108,116} { \fill[gray!70] ($(C)+(\a:3.4)$) circle (1pt); }
    \coordinate (ri)  at ($(C)+(100:3.4)$);
    \coordinate (Gj0) at (1.55,1.07);
    \coordinate (Ni)  at ($(ri)+(100:1.2)$);
    \draw[violet!12, line width=7pt] (Gj0) .. controls (1.75,1.85) and (2.4,2.3) .. (3.35,2.5);
    \draw[violet!70!black, line width=0.9pt] (Gj0) .. controls (1.75,1.85) and (2.4,2.3) .. (3.35,2.5)
        node[above left=-2pt, violet!70!black] {$\GG_j$};
    \draw[->, mygreen, line width=0.9pt] (ri) -- (Ni) node[left=-1pt, mygreen!80!black] {$\nn(\rr_i)$};
    \draw[->, dashed, orange!90!black, line width=0.9pt] (ri) -- (Gj0);
    \node[orange!90!black, anchor=west] at ($(ri)!0.5!(Gj0)+(0.05,-0.02)$) {$\|\GG_{j0}-\rr_i\|$};
    \pic[draw=orange!90!black, angle radius=5mm, angle eccentricity=1.6,
         pic text=$\theta$, pic text options={orange!90!black, font=\scriptsize}] {angle = Gj0--ri--Ni};
    \fill[black] (ri) circle (1.4pt) node[below=1pt] {$\rr_i$};
    \fill[violet!70!black] (Gj0) circle (1.4pt) node[right=1pt, violet!70!black] {$\GG_{j0}$};
    \node[anchor=north] at (1.9,-0.32)
        {$D=w^{\text{dist}}\,\|\GG_{j0}-\rr_i\|+w^{\text{ang}}\,(1-\cos\theta)$};
    \node at (1.9,-0.85) {(a)};
  \end{scope}
  \begin{scope}[xshift=4.9cm, yshift=2.15cm]
    \def\s{0.26}
    \node[anchor=south west, inner sep=0pt] at (-0.55,0.3) {$\XX\in\{0,1\}^{N^{\text{root}}\times N^{\text{guide}}}$};
    \foreach \j in {1,2,3} { \node[inner sep=0pt, font=\tiny] at ({(\j-0.5)*\s},0.14) {$\GG_{\j}$}; }
    \foreach \i in {1,3,6} { \fill[gray!14] (0,{-(\i-1)*\s}) rectangle ({3*\s},{-\i*\s}); }
    \foreach \i in {1,...,6} { \node[anchor=east] at (-0.06,{-(\i-0.5)*\s}) {$\rr_{\i}$}; }
    \foreach \i in {1,...,6} \foreach \j in {1,2,3} {
        \draw[gray!60, line width=0.4pt] ({(\j-1)*\s},{-(\i-1)*\s}) rectangle ({\j*\s},{-\i*\s}); }
    \foreach \i/\j in {2/1, 4/2, 5/3} {
        \fill[teal!70] ({(\j-1)*\s+0.03},{-(\i-1)*\s-0.03}) rectangle ({\j*\s-0.03},{-\i*\s+0.03});
        \node[white, font=\tiny] at ({(\j-0.5)*\s},{-(\i-0.5)*\s}) {1}; }
    \draw[decorate, decoration={brace, amplitude=3pt}] ({3*\s},{-6*\s-0.07}) -- (0,{-6*\s-0.07});
    \node[anchor=north, align=center] at ({1.5*\s},{-6*\s-0.2})
        {$\one^{\!\top}\XX=\one^{\!\top}$\\ one root per guide};
    \draw[decorate, decoration={brace, amplitude=3pt}] ({3*\s+0.07},0) -- ({3*\s+0.07},{-6*\s});
    \node[anchor=west, align=left] at ({3*\s+0.22},{-3*\s})
        {$\XX\one\le\one$\\ at most one\\ guide per root};
    \node[anchor=north east, gray!90, inner sep=0pt] at (3.2,{-6*\s-0.9}) {gray rows: unassigned roots $\mathcal{R}^{\text{free}}$};
    \node at (1.2,-3.0) {(b)};
  \end{scope}
\end{tikzpicture}
    \caption{\textbf{Binding cost and assignment constraints.} (a) The cost of Eq.~\eqref{eq:dgeo} combines the distance from the guide's scalp-side endpoint $\GG_{j0}$ to the root candidate $\rr_i$ with the angle $\theta$ between their connection and the scalp normal $\nn(\rr_i)$. (b) In the assignment matrix $\XX$ of Eq.~\eqref{eq:assign_ip}, each column has exactly one entry and each row at most one; empty rows are the unassigned roots $\mathcal{R}^{\text{free}}$.}
    \Description{}
    \label{fig:binding_geometry}
\end{figure}

\paragraph{Initial Assignment.}
Let $\XX \in \{0,1\}^{N^{\text{root}} \times N^{\text{guide}}}$ be the assignment matrix, where $\XX_{ij} = 1$ if and only if guide $\GG_j$ is assigned to root candidate $\rr_i$. We define a binding cost function $D$ that measures both the Euclidean distance from the guide's scalp-side endpoint to the candidate root, and the angular consistency between the connection vector and the scalp normal $\nn(\rr_i)$:
\begin{equation}\label{eq:dgeo}
\begin{aligned}
D(\rr_i, \GG_j) =
    w^{\text{dist}}\,\|\GG_{j0} - \rr_i\| + w^{\text{ang}} \left( 1 - \nn(\rr_i) \cdot \tfrac{\GG_{j0} - \rr_i}{\|\GG_{j0} - \rr_i\|} \right),
\end{aligned}
\end{equation}
where $\GG_{j0}$ denotes the scalp-side endpoint of guide $\GG_j$ and $w^\bullet$ are weights.
\autoref{fig:binding_geometry} (a) illustrates the two quantities.
Collecting all pairwise costs into a matrix $\CC \in \mathbb{R}^{N^{\text{root}} \times N^{\text{guide}}}$ with $\CC_{ij} = D(\rr_i, \GG_j)$, the guide-root binding can be formulated as the following integer programming problem:
\begin{equation}\label{eq:assign_ip}
\XX^\star = \arg\min_{\XX \in \{0,1\}^{N^{\text{root}} \times N^{\text{guide}}}} \;\langle \CC, \XX \rangle_F \quad \text{s.t.}~
\one^\top \XX = \one^\top, \XX \one \le \one.
\end{equation}
Here, $\langle \cdot, \cdot \rangle_F$ denotes the Frobenius inner product, and $\one$ denotes an all-ones vector of compatible dimension. The two constraints enforce that each guide is assigned to exactly one root candidate, while each root candidate is assigned to at most one guide. \autoref{fig:binding_geometry} (b) visualizes both constraints on $\XX$.
 Eq.~\eqref{eq:assign_ip} is a rectangular linear sum assignment problem, which can be solved using the Hungarian algorithm~\cite{Kuhn95}. The cost matrix $\CC$ is constructed in a single vectorized pass. The resulting initial root-guide correspondence defines a map $\Pi^{(0)}: \mathcal{G} \rightarrow \mathcal{R}$.

\paragraph{Parting-aware Re-assignment.}
Many hairstyles exhibit a clearly visible parting. However, under the cost function $D$ alone, which is purely geometric and does not account for the local coherence of the strand flow, a guide can be assigned to the wrong side of a parting, leading to undesirable crossing artifacts.

To extract the hair parting curve, we first grow a short one-segment strand from every assigned root in $\Pi^{(0)}(\mathcal{G})$ and project it onto the scalp surface $\mathcal{S}$, yielding a projected tangent vector $\tt_i^{\text{root}}$ for each root. We then compute the Voronoi adjacency $\mathcal{E}$ on $\mathcal{S}$, using the assigned roots $\Pi^{(0)}(\mathcal{G})$ as seeds for the partition. An edge $(\rr_i,\rr_j)$ is marked as \emph{opposing} if the two roots are adjacent under $\mathcal{E}$ and their projected tangent directions satisfy $\tt_i^{\text{root}} \cdot \tt_j^{\text{root}} < \tau$, where $\tau$ is a prescribed threshold $\tau = \cos(3\pi/4)$. We then collect the midpoints of all opposing root pairs and fit smooth curves on the scalp through these points. The final output is a set of partition curves $\Gamma = \{\gamma^{\text{part}}\}$, where each $\gamma^{\text{part}}$ divides its local scalp neighborhood into two sides labeled by $\sigma(\rr) \in \{+1,-1\}$.

\begin{figure}[t!]
    \centering
    \includegraphics[width=\linewidth]{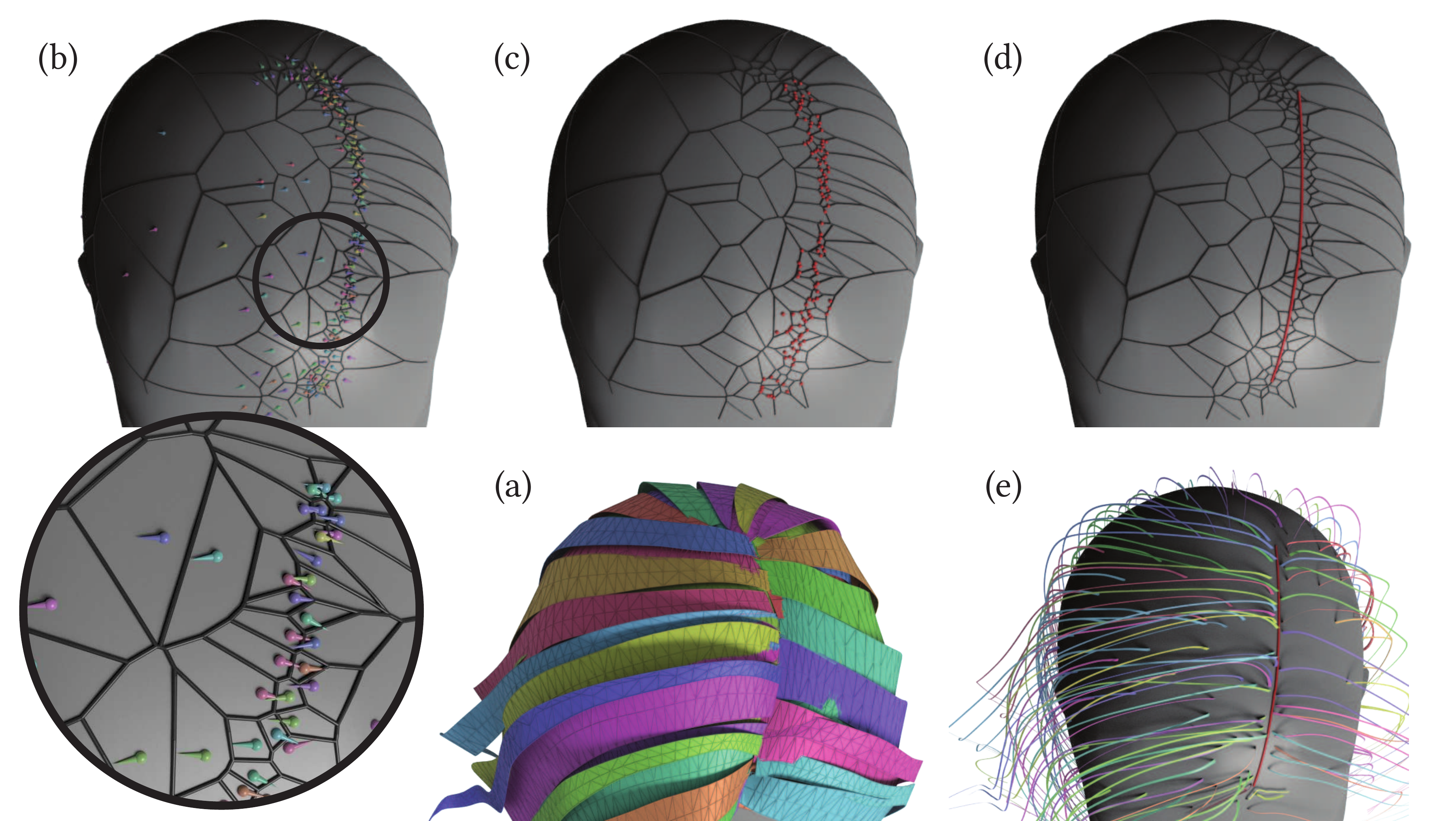}
    \caption{\textbf{Hair parting.} 
    Given the input hair cards and their extracted centerlines (a), each centerline is bound to the scalp and inherits an orientation vector projected onto the surface (b). We then construct a Voronoi tessellation on the scalp seeded by the bound roots, and identify candidate partition edges as those whose two adjacent cells carry opposing orientations. The midpoints of these edges are marked as candidate points (red dots in (c)). We fit an on-scalp curve through these candidates to obtain the final partition line (d). Finally, any guide strand that crosses the partition is reassigned to its correct side (e), yielding the consistent parting visible in the resulting guides.}
    \Description{}
    \label{fig:partition}
\end{figure}

We then refine the guide-to-root assignment using these partition curves. For an assigned root $\rr_i = \Pi^{(0)}(\GG_j)$ that lies within the influence region of some partition curve $\gamma^{\text{part}}$, we determine the side preferred by the guide's local flow as
\begin{equation}
\sigma^\star(\rr_i) = \operatorname{sign}\Bigl(\tt_i^{\text{root}} \cdot \bigl(\tt^{\text{part}}(\rr_i) \times \nn(\rr_i)\bigr)\Bigr),
\end{equation}
where $\tt^{\text{part}}(\rr_i)$ denotes the local tangent of $\gamma^{\text{part}}$ at the point on the curve closest to $\rr_i$. If $\sigma(\rr_i) \neq \sigma^\star(\rr_i)$, then the assigned root lies on the opposite side of the partition from the one implied by its strand flow. In that case, we replace $\Pi^{(0)}(\GG_j)$ with the nearest unoccupied dense root candidate located on the correct side of the partition. \autoref{fig:partition} demonstrates the process. The re-assigned correspondence defines the final map $\Pi: \mathcal{G} \rightarrow \mathcal{R}$ from guides to their roots.

\subsection{Root Tracing}\label{sec:tracing}

\begin{figure} [ht!]
    \centering
    \includegraphics[width=\linewidth]{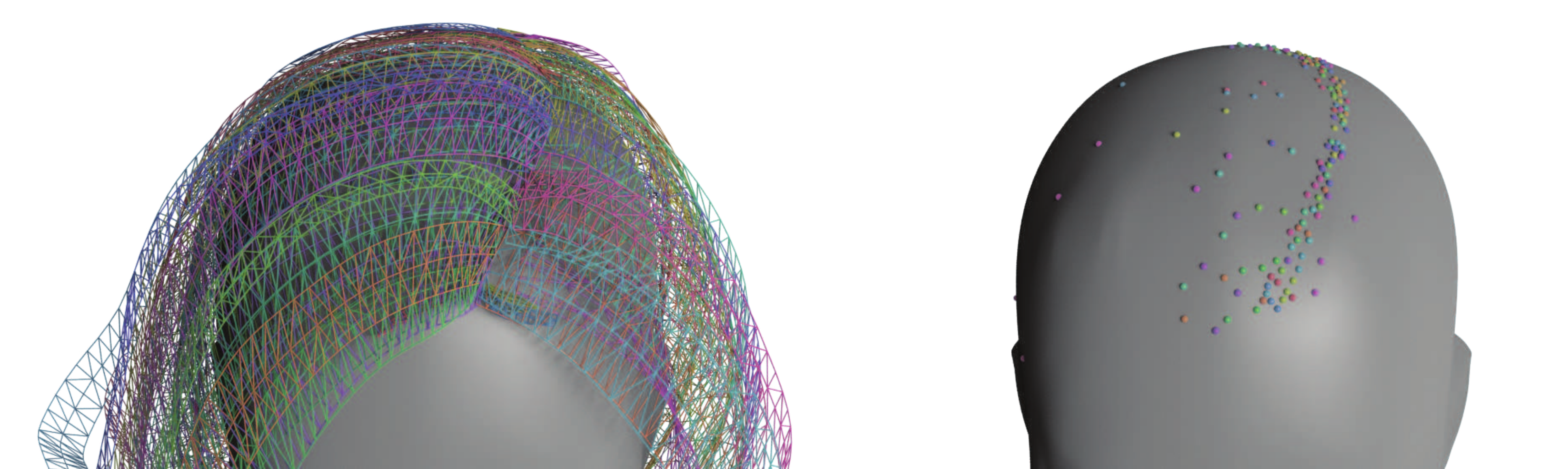}
    \caption{\textbf{Scalp coverage after binding.} Binding results solely depend on card placement, which may leave a large region of the scalp uncovered by the guides.}
    \Description{}
    \label{fig:tracing}
\end{figure}

The guide-to-root binding depends on the placement of the input hair cards. As a result, scalp regions that are far from any card endpoint may remain unsampled (see~\autoref{fig:tracing}), leaving gaps in root coverage. To fill these uncovered areas and support the following full-hair synthesis, we introduce a small number of additional guides.

\paragraph{Root Selection.}
Let $\mathcal{R}^{\text{free}} = \mathcal{R} \setminus$ $\Pi(\mathcal{G})$ denote the set of root candidates that have not been assigned. We select a budget of $N^{\text{extra}}$ additional roots from $\mathcal{R}^{\text{free}}$ using farthest-point sampling (FPS):
\begin{equation}\label{eq:fps}
    \mathcal{R}^{\text{extra}} \;=\; \operatorname{FPS}\bigl(\mathcal{R}^{\text{free}};\, N^{\text{extra}},\, \Pi(\mathcal{G})\bigr),
    \quad |\mathcal{R}^{\text{extra}}| = N^{\text{extra}}.
\end{equation}
The third argument initializes the FPS distance field with the existing card-seeded anchors. Consequently, each newly selected root is the point in $\mathcal{R}^{\text{free}}$ farthest from both the current anchor set and the roots selected in previous FPS iterations. In this way, the additional roots are distributed to cover the most under-sampled scalp regions. To avoid ambiguous assignments near hair partings, we exclude from $\mathcal{R}^{\text{free}}$ all roots that lie within a prescribed influence radius of any partition curve $\gamma^{\text{part}}$ $\in \Gamma$ before performing FPS.

\begin{figure}[t!]
    \centering
\begin{tikzpicture}[black, yscale=0.72, every node/.style={font=\scriptsize, inner sep=1.5pt}, >=stealth, line cap=round, line join=round]
  \begin{scope}
    \coordinate (C) at (2.6,-4.0);
    \fill[gray!10] ($(C)+(60:4.4)$) arc[start angle=60, end angle=120, radius=4.4] -- cycle;
    \draw[gray!60, line width=0.8pt] ($(C)+(60:4.4)$) arc[start angle=60, end angle=120, radius=4.4];
    \foreach \a in {64,72,94,106,118} { \fill[gray!60] ($(C)+(\a:4.4)$) circle (0.8pt); }
    \coordinate (A)  at ($(C)+(112:4.4)$);
    \coordinate (r1) at ($(C)+(100:4.4)$);
    \coordinate (r2) at ($(C)+(88:4.4)$);
    \coordinate (r3) at ($(C)+(76:4.4)$);
    \coordinate (r4) at ($(C)+(66:4.4)$);
    \coordinate (H1) at ($(r1)+(100:1.05)$);
    \coordinate (H2) at ($(r2)+(88:1.45)$);
    \coordinate (H3) at ($(r3)+(76:1.7)$);
    \coordinate (S)  at (1.0,0.85);
    \coordinate (E)  at (4.7,2.2);
    \draw[violet!12, line width=7pt] plot[smooth, tension=0.6] coordinates {(S) (H1) (H2) (H3) (E)};
    \draw[violet!70!black, line width=0.9pt] plot[smooth, tension=0.6] coordinates {(A) (S) (H1) (H2) (H3) (E)};
    \node[violet!70!black, anchor=south east] at ($(E)+(0.05,0.08)$) {$\CC_j$};
    \node[violet!70!black, anchor=east] at ($(S)+(-0.06,0.02)$) {$\GG_j$};
    \foreach \r/\h in {r1/H1, r2/H2, r3/H3} { \draw[->, mygreen, densely dotted, line width=0.6pt] (\r) -- (\h); }
    \draw[->, gray!70, dashed, line width=0.6pt] (r4) -- ($(r4)+(66:1.4)$) node[gray!80, anchor=south, inner sep=1pt] {no hit};
    \draw[teal!70!black, line width=0.8pt] plot[smooth, tension=0.5] coordinates {(r1) ($(H1)+(0,-0.20)$) ($(H2)+(0,-0.14)$) ($(H3)+(0,-0.07)$) (E)};
    \draw[teal!70!black, line width=0.8pt] plot[smooth, tension=0.5] coordinates {(r2) ($(H2)+(0,-0.42)$) ($(H3)+(0,-0.22)$) (E)};
    \draw[teal!70!black, line width=0.8pt] plot[smooth, tension=0.5] coordinates {(r3) ($(H3)+(0,-0.62)$) (4.4,1.9) (4.6,2.1) (E)};
    \node[teal!70!black] at (1.33,0.62) {$\mathcal{G}^{\text{extra}}$};
    \draw[<->, orange!90!black, line width=0.6pt] ($(H2)+(0.2,0)$) -- ($(H2)+(0.2,-0.42)$)
        node[midway, right=0pt, orange!90!black] {$o_i$};
    \draw[<->, gray!80, line width=0.5pt] ($(C)+(112:3.72)$) arc[start angle=112, end angle=88, radius=3.72];
    \node[gray!90] at ($(C)+(100:3.42)$) {$d_{i,j}$};
    \fill[violet!70!black] (A) circle (1.3pt) node[below=1pt, violet!70!black] {$\Pi(\GG_j)$};
    \fill[black] (r1) circle (1.3pt);
    \fill[black] (r2) circle (1.3pt) node[below=1pt] {$\rr_i$};
    \fill[black] (r3) circle (1.3pt) node[below=1pt, gray!90] {farthest};
    \fill[gray!70] (r4) circle (1.3pt);
    \node at (2.6,-0.9) {(a)};
  \end{scope}
  \begin{scope}[xshift=5.9cm, yshift=0.15cm]
    \draw[gray!60, dashed, line width=0.4pt] (1.7,0) -- (1.7,1.45) -- (0,1.45);
    \draw[->] (0,0) -- (2.15,0) node[below, inner sep=1pt] {$d_{i,j}$};
    \draw[->] (0,0) -- (0,1.8) node[left, inner sep=1pt] {$o_i$};
    \draw[orange!90!black, line width=0.9pt] (0,0) -- (1.7,1.45);
    \foreach \x/\y in {0.63/0.537, 1.27/1.083, 1.7/1.45} { \fill[teal!70!black] (\x,\y) circle (1.3pt); }
    \node[anchor=north] at (1.7,-0.03) {$D_j$};
    \node[anchor=east] at (-0.03,1.45) {$\epsilon_\text{offset}$};
    \node[anchor=north east, inner sep=1pt] at (0,-0.03) {$0$};
    \node at (1.0,-1.05) {(b)};
  \end{scope}
\end{tikzpicture}
    \caption{\textbf{Layered offset of traced guides.} (a) A root $\rr_i$ hosted by card $\CC_j$ is traced along the card and pushed toward the scalp by $o_i$, which grows with the scalp distance $d_{i,j}$ to the card's root and fades to zero at the tip; a root whose normal ray misses every card receives no guide. (b) $o_i$ rises linearly from $0$ to $\epsilon_\text{offset}$, reached by the farthest hosted root at $D_j$.}
    \Description{}
    \label{fig:layer_offset}
\end{figure}
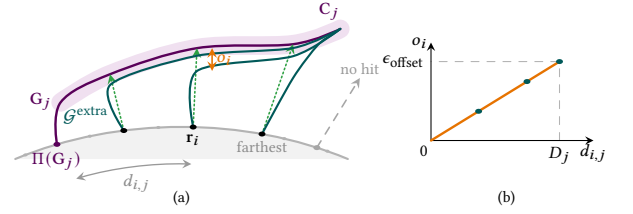

\paragraph{Tracing.}
For each $\rr_i \in \mathcal{R}^{\text{extra}}$, we shoot a ray from $\rr_i$ along the scalp normal $\nn(\rr_i)$ to the card set $\{\CC_j\}$. If it hits a card $\CC_j$, we trace a guide from the hit point along the flow direction to the tip, as determined in~\autoref {sec:extraction}. Such a root $\rr_i$ is hosted by card $j$ (\autoref{fig:layer_offset}). If the ray misses every card, no guide is generated from $\rr_i$.
Guides traced from the same card all lie on its surface and would collapse into a single flat sheet, even though the hair they represent fills the space between the card and the scalp. We therefore push each traced guide inward, toward the scalp, by an offset that grows with the distance of its root from the card's own root.  Concretely, let $d_{i,j} = \|\Pi(\GG_j) - \rr_i\|$ be the scalp distance from $\rr_i$ to the root of card $j$'s guide and $D_j = \max_i d_{i,j}$ its maximum over the roots hosted by card $j$. Every sample of the traced guide except the root is displaced along the inward card normal by $o_i = \epsilon_\text{offset}\, d_{i,j}/D_j$, scaled by a linear fade from one at the root end to zero at the tip so that the tip returns to the card surface. Roots next to the card's root keep their guides on the card, while the farthest hosted root attains the full $\epsilon_\text{offset}$ and runs closest to the scalp, which yields the layered placement of \autoref{fig:layer_offset}.
The resulting traced guides from $\mathcal{R}^{\text{extra}}$ form an extra guide set $\mathcal{G}^{\text{extra}}$. The full guide set is $\mathcal{G}^{\star} = \mathcal{G} \cup \mathcal{G}^{\text{extra}}$, and $\Pi$ extends to $\mathcal{G}^{\star}$ by mapping each traced guide to the root in $\mathcal{R}^{\text{extra}}$ it was traced from.

\begin{figure}[t!]
    \centering
    \includegraphics[width=\linewidth]{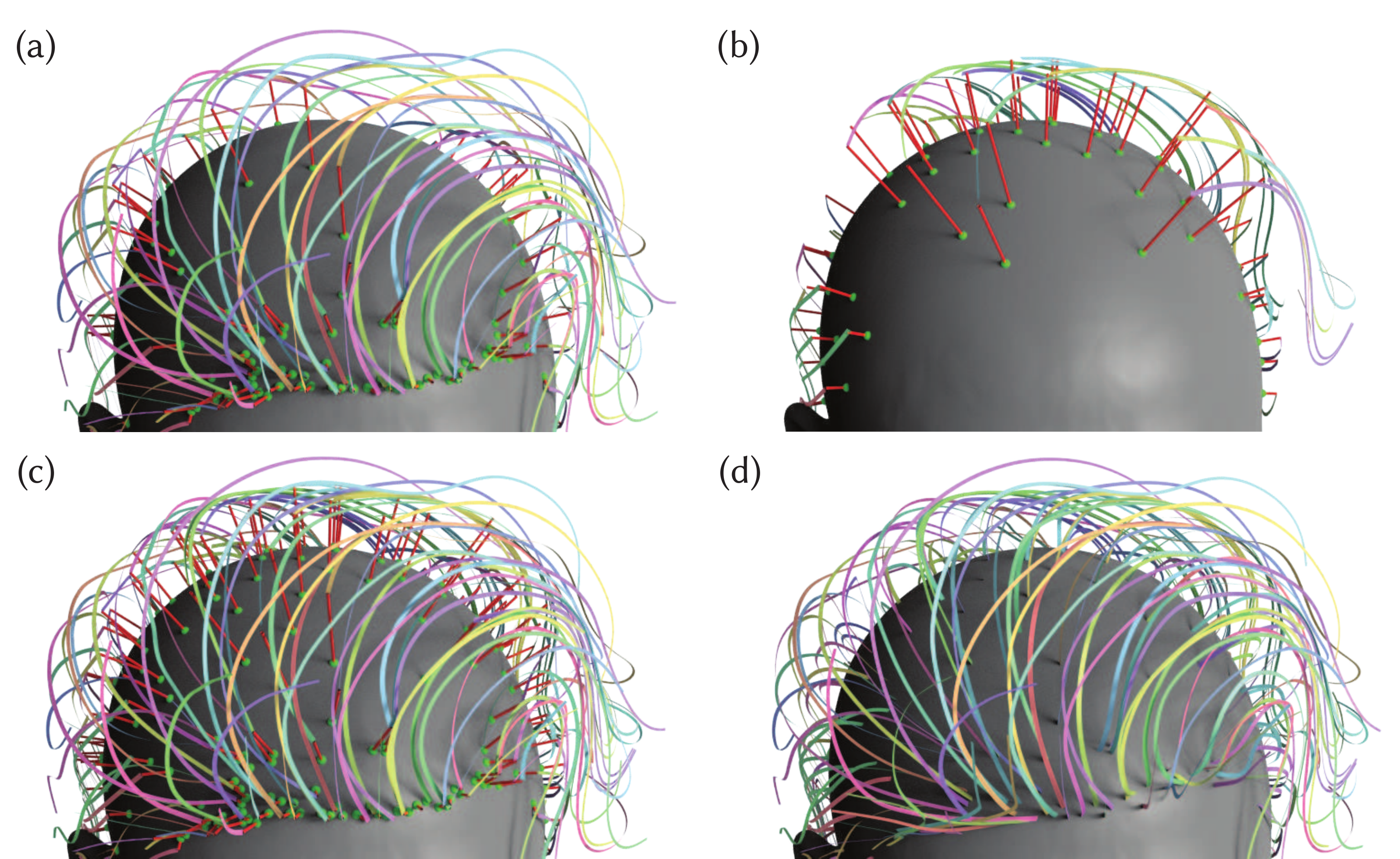}
    \caption{\textbf{Guide generation.} We begin by binding the centerline of each card to form an initial set of guide strands (a). We then sample a sparser set of unbound root candidates, trace each one along the scalp's normal direction until it hits the nearest card, and continue the remainder of the strand along that card's orientation (b). The resulting guide set $\mathcal{G}^\star=\mathcal{G}\cup\mathcal{G}^\text{extra}$ shown in (c) is then smoothed near the roots to reduce kinks, yielding the final guide output (d). }
    \label{fig:hair_tracing}
    \Description{}
\end{figure}

\paragraph{Smoothing.}
Because the guides may exhibit a visible kink at their scalp connections, smoothing is necessary near the roots. However, naive Laplacian smoothing of the guide samples tends to wash out the original input curvature and may even cause the guide to penetrate the scalp surface. We therefore formulate a geometric optimization problem that balances several objectives, including length preservation, root-orientation consistency, proximity to the original guide, and smoothness.
Let $\gg = (\gg_0, \dots, \gg_K)$ denote the root-region samples of a guide in $\mathcal{G}$, where $\gg_0$ is fixed at the scalp root and $\gg_K$ is fixed at the junction point offset from the on-card portion. Let $\gg^{(0)}$ denote the initial sample positions before smoothing. We optimize the free variables $(\gg_1, \dots, \gg_{K-1})$ by minimizing
\begin{equation}\label{eq:smooth}
\begin{aligned}
\mathcal{L}^{\text{smooth}}(\gg)
=\;&
w^{\text{len}} \bigl(L(\gg) - L(\gg^{(0)})\bigr)^2 + w^{\text{ori}} \bigl\|\bar{\ee}_1 -\nn(\gg_0)\bigr\|^2
\\
+ & ~w^{\text{proj}} \sum_{k=1}^{K-1} d^2\bigl(\gg_k,\, \gg^{(0)}\bigr) + w^{\text{lap}} \sum_{k=1}^{K-1} \|\Delta_k \gg\|^2 ,
\end{aligned}
\end{equation}
where $L(\gg) = \sum_k \|\gg_{k+1} - \gg_k\|$ is the total arc length, $\bar{\ee}_1 = (\gg_1 - \gg_0) / ||\gg_1 - \gg_0||$ is the unit vector of the first segment, $d(\pp, \gg^{(0)})$ denotes the projection distance from a point $\pp$ to $\gg^{(0)}$, and $\Delta_k \gg = \gg_{k-1} - 2\gg_k + \gg_{k+1}$ is the discrete Laplacian. The second term penalizes deviation of the first segment from the scalp normal direction $\nn(\gg_0)$, encouraging the guide to emerge naturally from the scalp. Eq.~\eqref{eq:smooth} is solved independently for each guide in parallel using fixed-step gradient descent. During each iteration, the free sample indices are updated in root-to-tip order using a standard Gauss-Seidel sweep. The whole process is illustrated in~\autoref{fig:hair_tracing}.

\subsection{Wrapper Relaxation}\label{sec:relaxation}
Given the full guide set $\mathcal{G}^{\star}$ generated in the previous stages, the guides are either aligned with the input hair cards or extended from scalp roots in directions parallel to the cards. However, their spatial distribution and spacing are generally uneven. In a standard hair grooming pipeline, each guide typically represents a cluster of individual hair strands. Accordingly, each guide can be viewed as occupying a volumetric region that originates from an associated area on the scalp, which we refer to as a \emph{wrapper}, denoted as $\{\mathcal{W}_j\}$. Each wrapper is a tube-shaped volume. The size of this area is related to the number of physical strands represented by the guide. If the guides are distributed unevenly, the synthesized hair may exhibit undesirable artifacts, such as visible gaps between strand clusters or excessive crowding in a confined region. Neither effect is desirable for downstream hair generation.
To regularize the guides and adjust their relative positions, we formulate a wrapper relaxation step. The goal is to obtain a well-distributed guide set whose effective guide volumes, represented by the wrappers, do not overlap excessively, while still preserving the initial structure and overall silhouette specified by the input hair cards.

\paragraph{Wrapper Initialization.}
We first compute the geodesic Voronoi diagram of the root set $\{\Pi(\GG_j)\}$ over the scalp surface $\mathcal{S}$, and then clip the resulting cells by the partition-curve set $\Gamma$. The boundary of each clipped Voronoi cell is taken as the cross-section of the corresponding wrapper at the scalp. By construction, these Voronoi cells form a partition of $\mathcal{S}$ without gaps or overlaps, and any two adjacent cells share a common boundary curve. For each guide, we map the corresponding cell boundary back to its 3D positions on the scalp and propagate it along the guide using the parallel-transport frame of $\GG_j$ to construct a target ring at each layer for relaxation. Each cross-layer is initialized as a thin spindle to avoid initial overlap, by rescaling the target ring's radius about the centerline.

\paragraph{SDF Construction.}
We further construct a hair volume $\Omega^{\text{card}}$ on a voxel grid with resolution $h_\Omega$ by rasterizing the input card set $\{\CC_i\}$ onto the grid using a thickness parameter $\epsilon^{\text{thickness}}$. We then identify the interior void between the hair volume and the bust mesh as $\Omega^{\text{gap}}$. The total admissible volume is defined as the union $\Omega = \Omega^{\text{card}} \cup \Omega^{\text{gap}}$. From this volume, we build a signed distance field $\Phi_\Omega$, where $\Phi_\Omega \le 0$ inside the volume and $\Phi_\Omega > 0$ outside, using an unsigned distance transform together with an inside-outside classification test. We also construct a separate signed distance field $\Phi_{\mathcal{B}}$ for the bust geometry to support collision handling. To ensure stable, well-behaved gradients during relaxation, we apply a Gaussian blur to both fields.
Bare scalp regions contain no cards, so $\Omega$ excludes them and no wrapper is placed there. We extract the boundary of the covered scalp region as one or several closed curves, and the enclosed region serves as the bottom of the wrappers.

\paragraph{Wrapper Relaxation.}
Each wrapper is initialized as a thin spindle and then expanded during relaxation, allowing it to gradually occupy volume and separate from neighboring wrappers in a controlled manner. At each iteration, all ring vertices are expanded with a fixed step along the outward cross-sectional direction. A vertex stops expanding once its distance from the centerline reaches $\lambda_{\text{cap}}$ times its target-ring radius, which keeps a wrapper that faces empty space close to its scalp footprint. The proposed per-vertex displacement is then clamped using edge-edge additive continuous collision detection (ACCD)~\cite{Li2021CIPC} against all wrapper edges, as well as against the hair-volume boundary $\partial\Omega$. After clamping, Laplacian smoothing is applied along the longitudinal direction of each wrapper to remove the local zig-zag artifacts introduced by collision handling.

\begin{figure}
    \centering
    \includegraphics[width=1.0\linewidth]{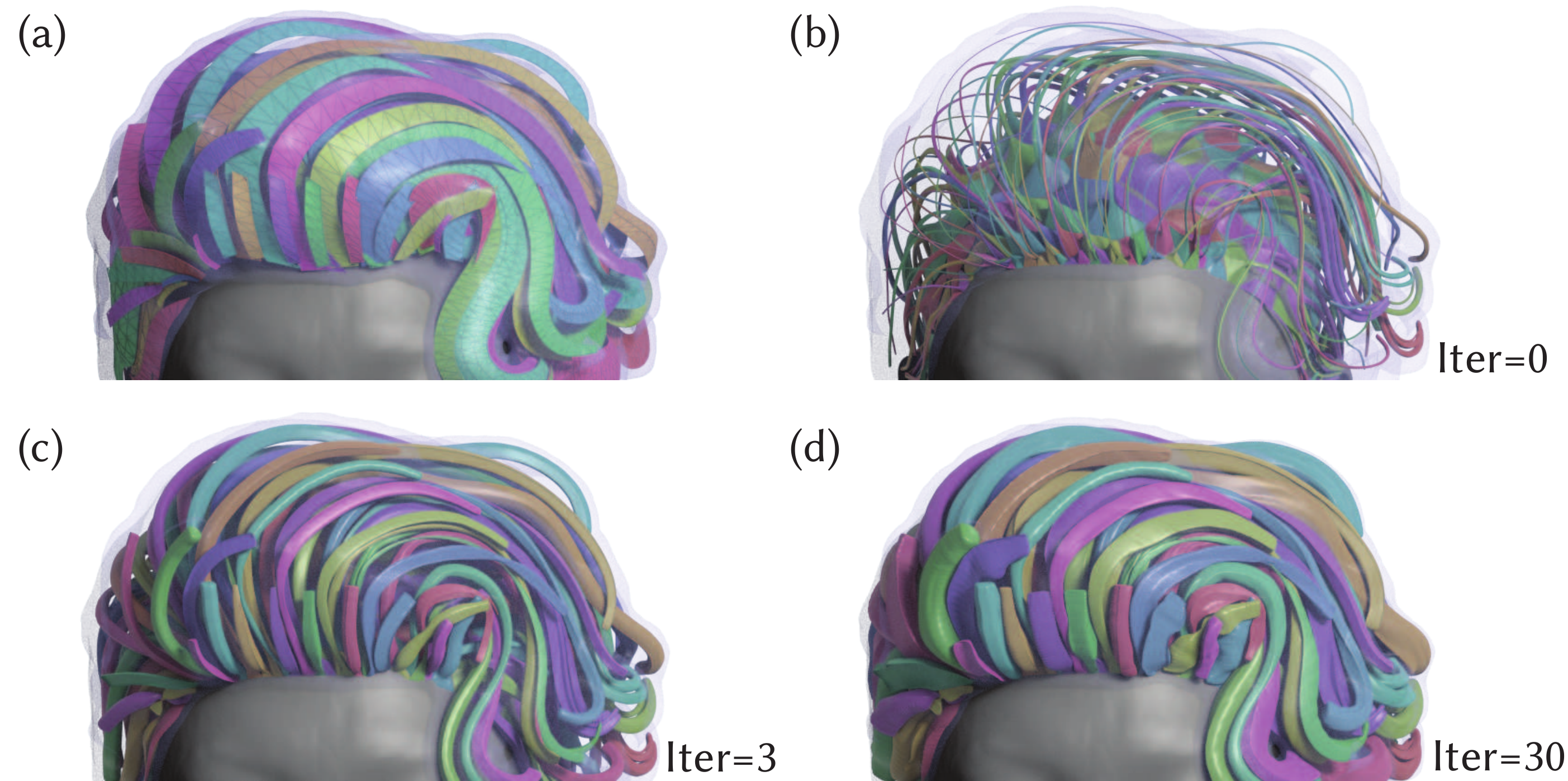}
    \caption{\textbf{Wrapper relaxation.} We construct the SDF of the hair volume $\Omega$, which encloses the thickened cards and the space between the hair cards and the bust (a). Each guide's wrapper is initialized as a thin spindle that partitions the scalp region (b). After relaxation (c-d), every wrapper represents the volumetric occupancy of its wisp, and together they form a non-overlapping partition of the whole volume of the hairstyle. }
    \Description{}
    \label{fig:wrapper_relaxation}
\end{figure}

In this manner, each wrapper grows progressively over multiple iterations. Expansion continues until ACCD reduces the displacement of every ring vertex to zero, indicating that the wrapper has fully occupied the available space between neighboring wrappers, the outer hair-volume boundary $\partial\Omega$, and the bust geometry $\mathcal{B}$. 
We model the guides as elastic rods driven by the collision penalty against the wrappers and solve for their deformation with an XPBD solver. As the wrappers expand, treating the guides as elastic rods, rather than displacing vertices independently, preserves each guide's original shape and curvature, admitting only mild non-smooth variation. Once the solver commits the displacement updates of the guide strands, a CCD pass follows to ensure that no two guides cross each other and that all remain within the admissible hair volume defined by $\Phi_\Omega$. Finally, we resynchronize the rings of each wrapper with the relaxed guide by updating them using the guide's parallel-transport frame and wrapper radius, and then perform a final CCD pass. \autoref{fig:wrapper_relaxation} illustrates the relaxation process and \autoref{apdx:relaxation} provides a detailed algorithm of wrapper relaxation. 

\subsection{Full Hair Synthesis}\label{sec:full-hair}
Given the relaxed guides and their associated wrappers, the final step is to synthesize the full set of hair strands. For each wrapper associated with the relaxed guide $\GG_i$, we first transport an orthonormal frame along $\GG_i$ using parallel transport. At each layer, this frame is then deformed within the cross-sectional plane to match the local shape of the wrapper. We realize this deformation by constructing a per-layer scale mapping from the transported frame to the corresponding wrapper cross-section in axis-angular coordinates. In this way, each synthesized strand inherits the tangent profile of $\GG_i$, is rooted at $\rr_i$ on the scalp, and remains inside the wrapper by construction (see~\autoref{apdx:wrapper_deformation}). However, this per-wrapper initialization is only locally consistent and remains globally piecewise. In particular, across wrapper boundaries, strand tangents may change abruptly because strands on the two sides are generated from different guides, leading to visible gaps or discontinuities. To address this issue, we perform an additional strand-level relaxation to smooth the strand distribution throughout the hair volume.

\begin{figure*}
    \centering
    \includegraphics[width=1.0\linewidth]{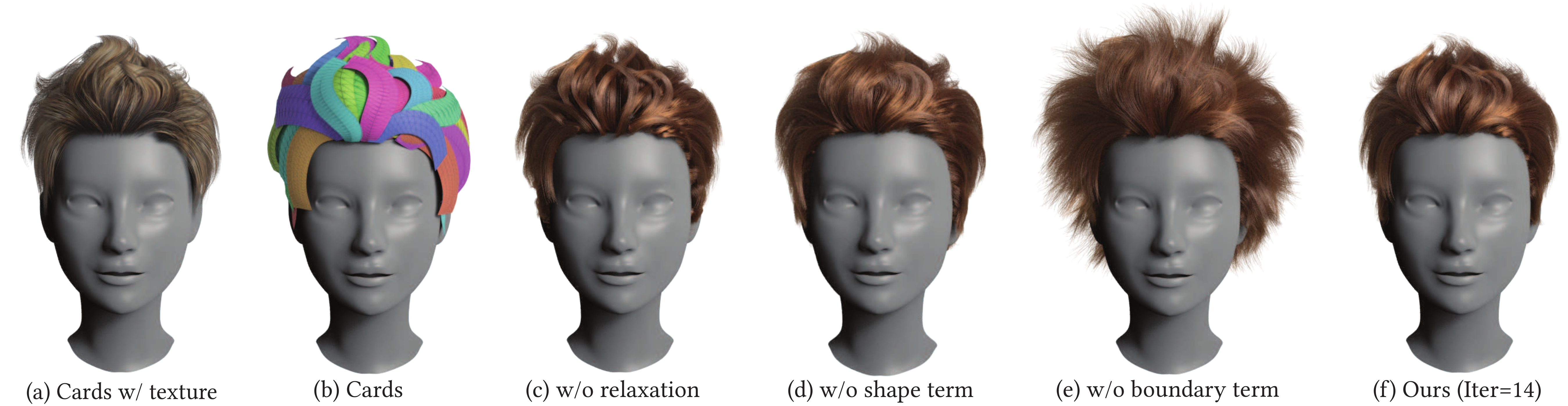}
    \caption{\textbf{Ablation studies on full-strand relaxation.} We show the effect of each term in the relaxation process after generating the initialized strands within the wrapper. (a) Rendered hair cards. (b) Input hair-card geometry. (c) Initialized result without relaxation. (d) Result without the shape regularization term. (e) Result without the wrapper and the bust boundary terms. (f) Result using the full energy after convergence.}
    \Description{}
    \label{fig:density}
\end{figure*}

Specifically, we represent each hair strand $\SS_i = \{\ss_{ij}\}$, with $N = |\SS_i|$ samples, as a chain of fixed-length segments connected by hinge joints, parameterized by $\QQ_i = \{\qq_{ij}\}$. The root vertex is pinned to the scalp, and each segment length is preserved at its initialized value. At every interior joint $j$, we introduce two rotational degrees of freedom,
$\qq_{ij} = (\theta_{ij}, \phi_{ij})$,
defined as in-plane rotation angles in the basis spanned by the two principal directions of the plane normal to the current segment. 
We optimize the per-strand configuration by minimizing
\begin{equation}\label{eq:density_energy}
\begin{aligned}
   \mathcal{L}^{\text{relax}}_i(\QQ_i)
   = \sum_{j=1}^{N-1} \sum_{\bullet\in\{\text{shape},\text{den},\text{wrap},\text{bust}\}} 
         w^{\bullet}\, \mathcal{L}^{\bullet}_j
      ,
\end{aligned}
\end{equation}
which balances shape preservation, density regularization, wrapper containment, and bust collision avoidance. $w^{\bullet}$ are the corresponding weights, and the individual terms are defined as
\begin{align}
     \mathcal{L}^{\text{shape}}_j
        &= w_j \bigl(1 - \tt_{ij} \cdot \tt^{\text{guide}}_{ij}\bigr), \quad\;\;
     \mathcal{L}^{\text{den}}_j
        = \bigl(\rho(\ss_{ij}) - \rho^\star_{\OO}(\ss_{ij})\bigr)^2, \label{eq:relaxation_terms} \\
     \mathcal{L}^{\text{wrap}}_j
        &= \max\bigl(0,\, \Phi_\Omega(\ss_{ij}) + \epsilon\bigr)^2, \;
     \mathcal{L}^{\text{bust}}_j
        = \max\bigl(0,\, \epsilon - \Phi_\mathcal{B}(\ss_{ij})\bigr)^2. \notag
\end{align}
The vector $\tt_{ij}$ is the current unit segment direction at joint $j$, and $\tt^{\text{guide}}_{ij}$ is the unit tangent at the corresponding vertex of $\GG_i$. The weight $w_j \in [0,1]$ is a root-to-tip ramp: it increases linearly from $0$ over the first few joints and reaches $1$ afterward. This design makes the strand near the scalp less constrained by the guide shape, while the tip region follows the guide tangent more strongly, helping preserve the layered structure.
The function $\rho(\xx)$ denotes the segment-rasterized strand density on the wrapper voxel grid at location $\xx$, and $\rho^\star_{\OO}(\xx)$ is the target density obtained by diffusing the scalp root density throughout $\Omega$ along a precomputed orientation field $\OO$ derived from the initialized strands. The wrapper and bust terms are formulated as one-sided SDF penalties with a small margin $\epsilon$.
We minimize $\mathcal{L}^{\text{relax}}_i$ using Adam-based~\cite{kingma2015adam} gradient descent over rotational variables $\QQ_i$, sweeping the joints from root to tip.

\section{Results}

\begin{figure}
    \centering
    \includegraphics[width=1.0\linewidth]{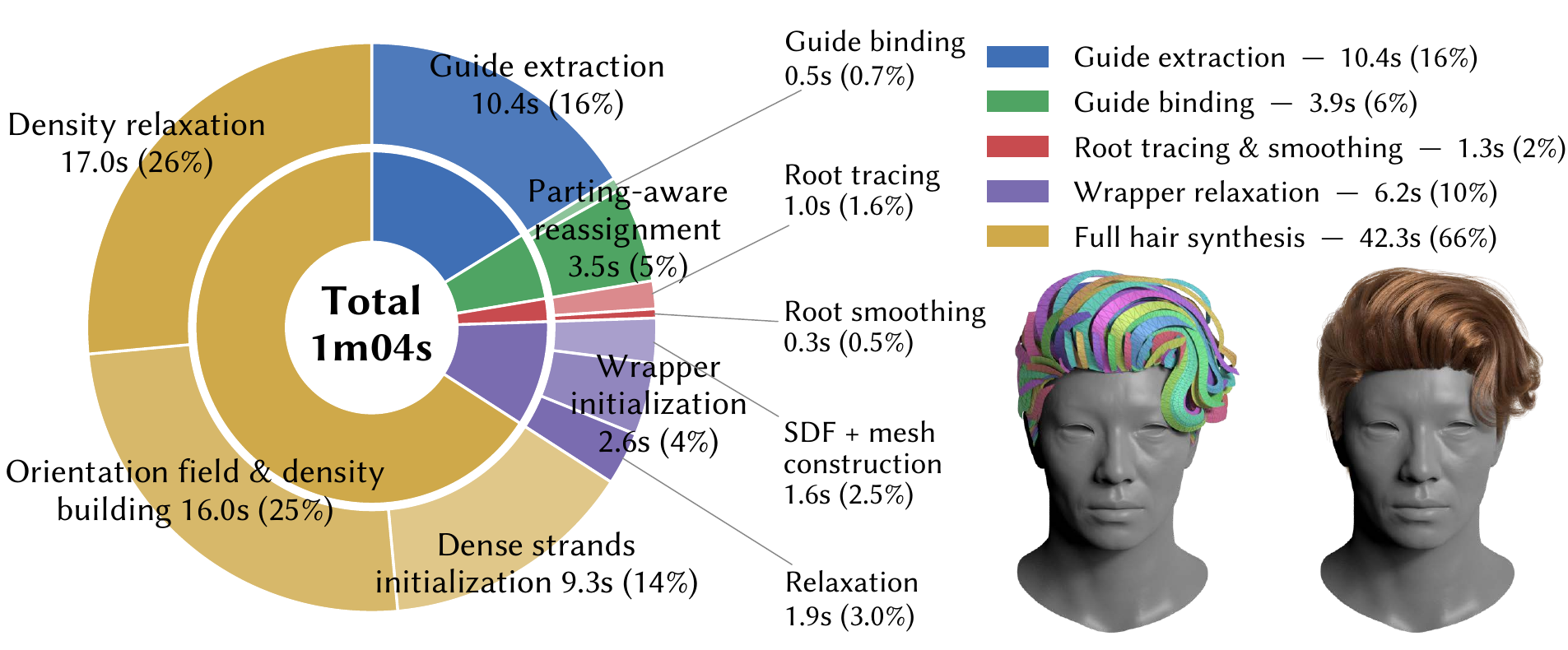}
    \caption{\textbf{Time Statistics.} Detailed time statistics for each component in our pipeline. The most time-consuming stage is full-hair synthesis, as it directly manipulates the dense strands. Our method typically runs for about $1{-}2$ minutes per case.
    \label{fig:time_memory}}
    \Description{}
\end{figure}

We implement our system on a desktop platform with an \textsf{AMD} \textsf{Ryzen} \textsf{9} \textsf{9950X} 16-core CPU, a GPU with 16,384 cores and 24GB of VRAM, and 64GB RAM. We implement root smoothing, wrapper relaxation, and hair synthesis on the GPU through the Nvidia Warp library~\cite{warp2022}. \autoref{fig:time_memory} shows the time statistics for each component of our pipeline. Our method runs in about $1{-}2$ minutes for about $100K$ dense strands on hundreds of input hair cards, and has a memory footprint of less than $500$MB.

Every result in this paper, including the in-the-wild assets of \autoref{sec:generalization} and the 50K dataset, is produced with the single parameter set listed in \autoref{tab:hparams}; no parameter is tuned per asset. The one documented exception is the sample count per guide and per strand, which is raised to 128 for the strongly curved in-the-wild cards, as stated in \autoref{sec:generalization}. In all experiments, we normalize the head mesh to unit scale. Guides and strands carry $|\SS_i| = 32$ samples, or $64$ when grooming operators are applied, all fields are computed on a $128^3$ grid, and the density relaxation runs the Adam optimizer until the loss drops to $10\%$ of its initial value; the weights, thresholds, and iteration counts of every stage are given in \autoref{tab:hparams}.

\begin{figure*}
    \centering
    \includegraphics[width=\linewidth]{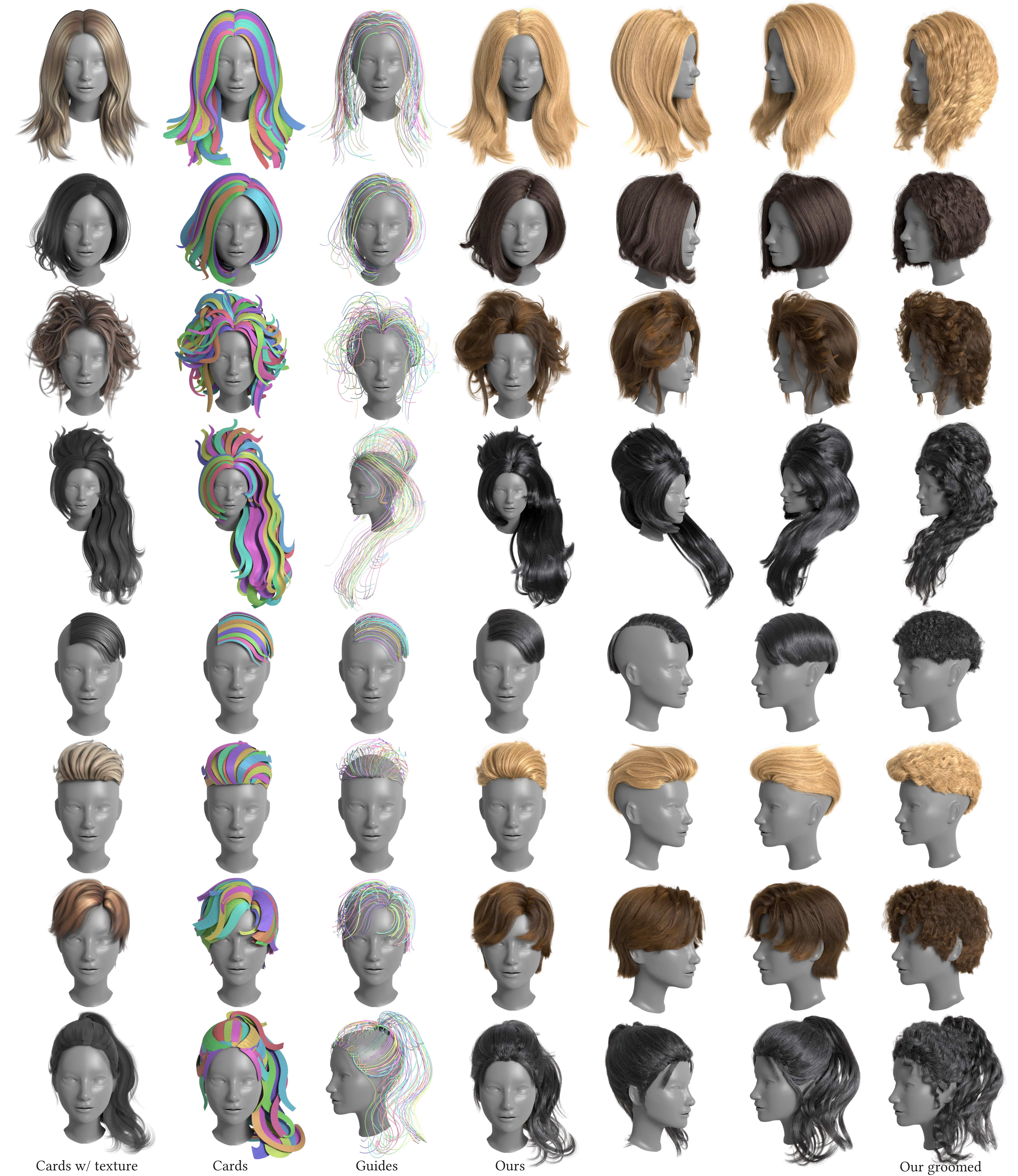}
    \caption{We evaluate our method on a variety of hairstyles from~\citet{Zheng2024}. Our approach robustly reproduces faithful results even for challenging wisp structures. The last column shows results obtained by using our strand-based model as the base hair, with grooming operators such as helix and fuzz to enrich its features. Details about helix and curly operators are provided in~\autoref{apdx:wrapper_deformation} and~\autoref{apdx:helix_operator}.
    }
    \Description{}
    \label{fig:evaluation}
\end{figure*}

\begin{figure*}[t]
    \centering
    \includegraphics[width=\linewidth]{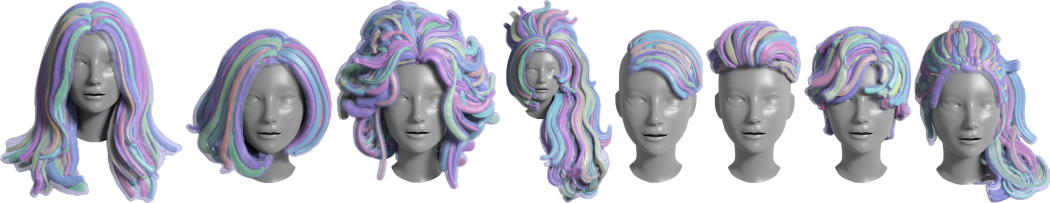}
    \caption{\textbf{Relaxed wrappers.} The relaxed wrappers of the examples in \autoref{fig:evaluation}. Each wisp wrapper is rendered in the color of its card, and the hair volume is rendered as a transparent blue shell.}
    \Description{}
    \label{fig:wrappers}
\end{figure*}

\begin{figure*}[p]
    \centering
    \includegraphics[width=\textwidth]{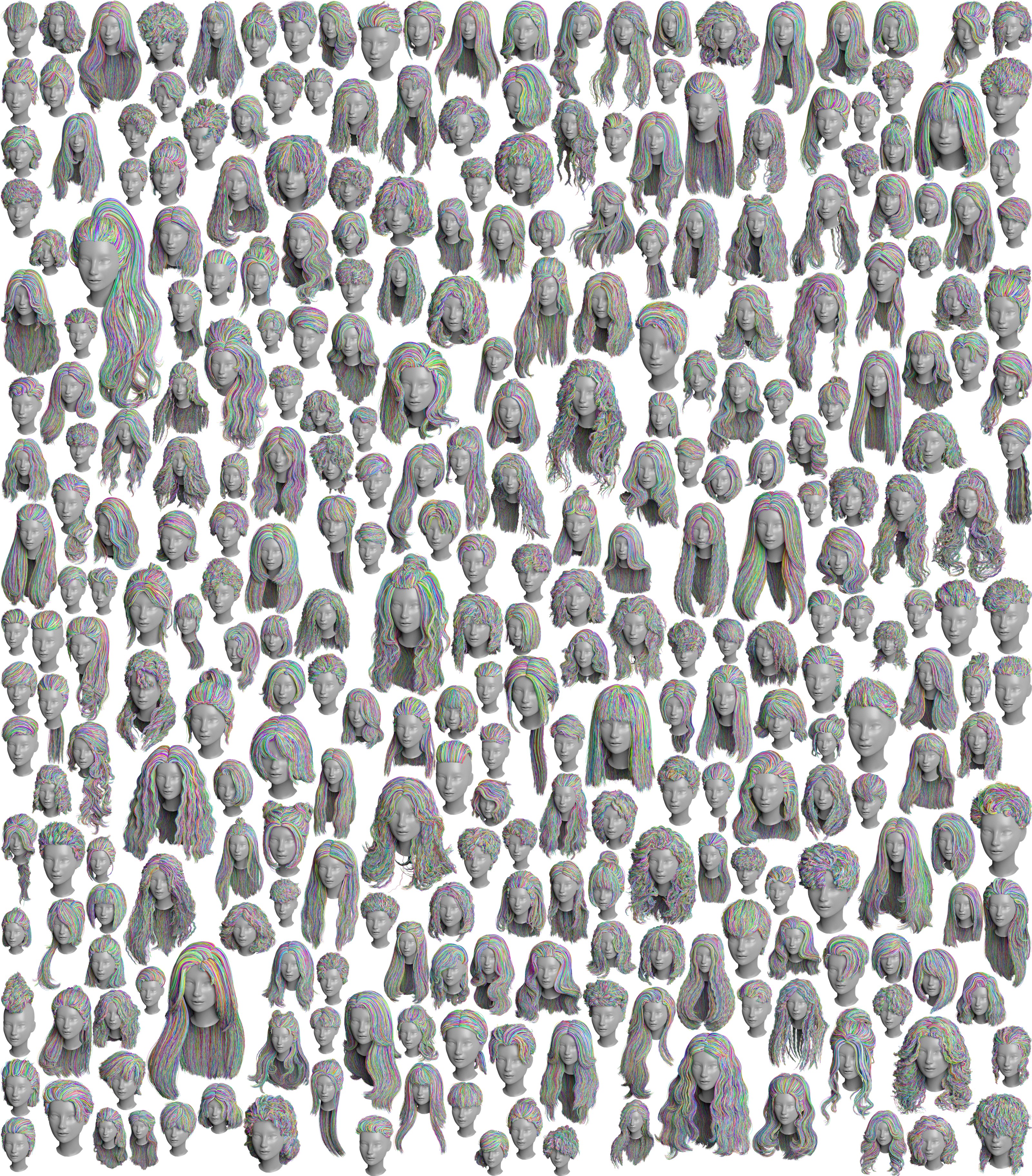}
    \caption{\textbf{Gallery.} 300 example strand-based hairstyles randomly chosen from our 50K dataset created by blending the strand-based models generated by our method.
    }
    \Description{}
    \label{fig:gallary}
\end{figure*}

We evaluate our method on a large-scale hair-card dataset covering a wide variety of hairstyles, including short, long, and wavy hair, as well as more complex styles such as ponytails and buns. We use hair cards from a large hairstyle dataset~\cite{Zheng2024}. These cards are artist-authored game assets collected from The Sims Resource. They ship with shared texture atlases in which one atlas island serves many cards, and their UV conventions are inconsistent: the hair flow runs along either the $u$- or the $v$-axis and the root lies at either end, often within the same asset. \autoref{fig:evaluation} shows a subset of the results, and \autoref{fig:wrappers} shows the relaxed wrappers from which their wisps are synthesized. In addition, we present a gallery of approximately three hundred strand-based hairstyles generated by our method (see \autoref{fig:gallary}).





\begin{figure} [t!]
    \centering
    \includegraphics[width=\linewidth]{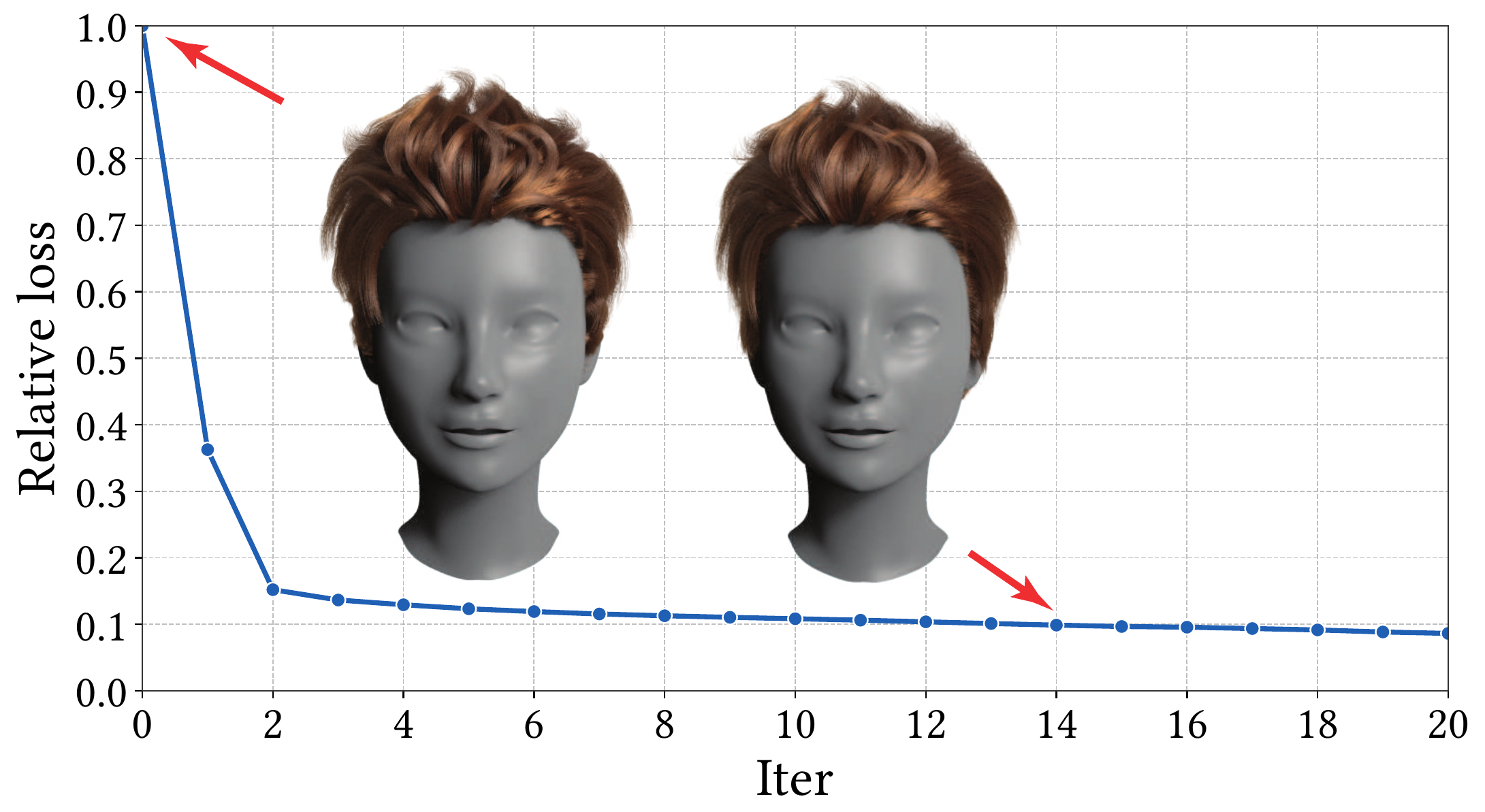}
    \caption{\textbf{Relaxation Loss.} We evaluate the relative loss with respect to the non-relaxed result. After convergence, the relaxation effectively removes the non-smooth boundaries between wisps.}
    \Description{}
    \label{fig:loss}
\end{figure}

\subsection{Evaluations}

\begin{table*}[ht!]
  \centering
  \caption{Evaluation on the representative models from the dataset~\cite{Zheng2024}, shown in~\autoref{fig:evaluation}. Here, $N_\text{card}$ denotes the number of hair-card components ($|\{\CC_i\}|$) in the input mesh, $N_\text{guide}$ denotes the number of guide strands ($|\{\mathcal{G}^\star\}|$), $N_\text{strand}$ denotes the number of final strands ($|\{\SS_i\}|$), and $N_\text{point}$ denotes the total number of strand vertices. IoU is the intersection-over-union of the card and strand hair silhouettes. PSNR and SSIM are reported separately for the tangent and the depth AOV renders of the hair cards (ground truth) and of our strands (\autoref{fig:tangent_eval}). Both are rendered from the same camera and compared on their shared non-transparent bounding box after compositing over a constant background. $d_{c\to s}$ is the mean distance from the opaque card surfaces to the nearest strand (Eq.~\eqref{eq:dcs}). CD denotes the symmetric Chamfer distance between the strand vertices and an equal-size uniform sample of the hair volume, with both point sets capped at 200K points.  Time reports the total wall-clock time of the full pipeline used to generate the strand-based results.}
  \label{tab:evaluation}
  \footnotesize
  \setlength{\tabcolsep}{2.2pt}
  \begin{tabular}{lrrrrrrrrrrrr}
    \toprule
    & \multicolumn{4}{c}{Strand statistics} & \multicolumn{6}{c}{Fidelity to cards} & \multicolumn{1}{c}{Distribution} & \multicolumn{1}{c}{Time} \\
    \cmidrule(lr){2-5} \cmidrule(lr){6-11} \cmidrule(lr){12-12} \cmidrule(lr){13-13}
    & & & & & \multicolumn{1}{c}{Silhouette} & \multicolumn{2}{c}{Tangent AOV} & \multicolumn{2}{c}{Depth AOV} & \multicolumn{1}{c}{Geometry} & & \\
    \cmidrule(lr){6-6} \cmidrule(lr){7-8} \cmidrule(lr){9-10} \cmidrule(lr){11-11}
    Examples & $N_\text{card}$ & $N_\text{guide}$ & $N_\text{strand}$ & $N_\text{point}$ & IoU$\uparrow$ & PSNR$\uparrow$ & SSIM$\uparrow$ & PSNR$\uparrow$ & SSIM$\uparrow$ & $d_{c\to s}\downarrow$ & CD$\downarrow$ & (min) \\
    \midrule
    row~1 & 62 & 208 & 83,893 & 2,684,576 & 0.838 & 16.76 & 0.562 & 15.00 & 0.523 & 0.00100 & 0.0052 & 1.6 \\
    row~2 & 50 & 218 & 83,825 & 2,682,400 & 0.895 & 18.07 & 0.574 & 15.73 & 0.517 & 0.00077 & 0.0047 & 1.1 \\
    row~3 & 122 & 284 & 90,245 & 2,887,840 & 0.882 & 19.04 & 0.562 & 16.92 & 0.521 & 0.00133 & 0.0051 & 1.2 \\
    row~4 & 92 & 267 & 105,605 & 3,379,360 & 0.860 & 16.84 & 0.531 & 15.93 & 0.512 & 0.00152 & 0.0063 & 1.2 \\
    row~5 & 29 & 201 & 47,365 & 1,515,680 & 0.963 & 19.20 & 0.755 & 17.93 & 0.743 & 0.00044 & 0.0031 & 0.7 \\
    row~6 & 44 & 218 & 58,342 & 1,866,944 & 0.966 & 21.49 & 0.733 & 20.53 & 0.726 & 0.00085 & 0.0038 & 0.7 \\
    row~7 & 86 & 243 & 81,345 & 2,603,040 & 0.948 & 19.65 & 0.632 & 18.07 & 0.590 & 0.00067 & 0.0040 & 1.4 \\
    row~8 & 110 & 308 & 82,966 & 2,654,912 & 0.889 & 17.83 & 0.579 & 17.39 & 0.559 & 0.00134 & 0.0054 & 1.4 \\
    \bottomrule
  \end{tabular}
\end{table*}

\begin{figure*}[t]
    \centering
    \includegraphics[width=\linewidth]{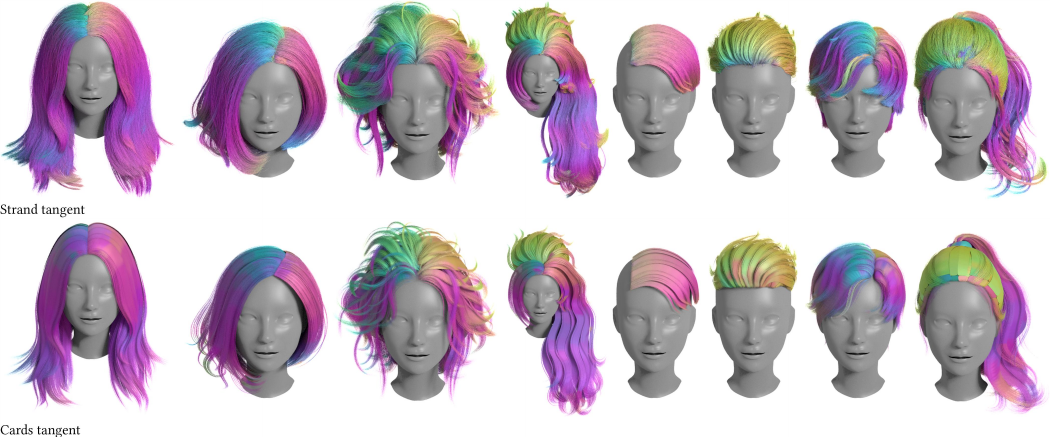}
    \caption{\textbf{Tangent renders for the image-space metrics.} Per-pixel tangent of our strands (top) and of the input cards (bottom) for the eight examples of \autoref{fig:evaluation}, rendered from the same camera. Encoding the tangent as color exposes the wisp structure and parting lines while removing the shading difference between cards and strands.}
    \Description{}
    \label{fig:tangent_eval}
\end{figure*}

\begin{figure}[t]
    \centering
    \includegraphics[width=\linewidth]{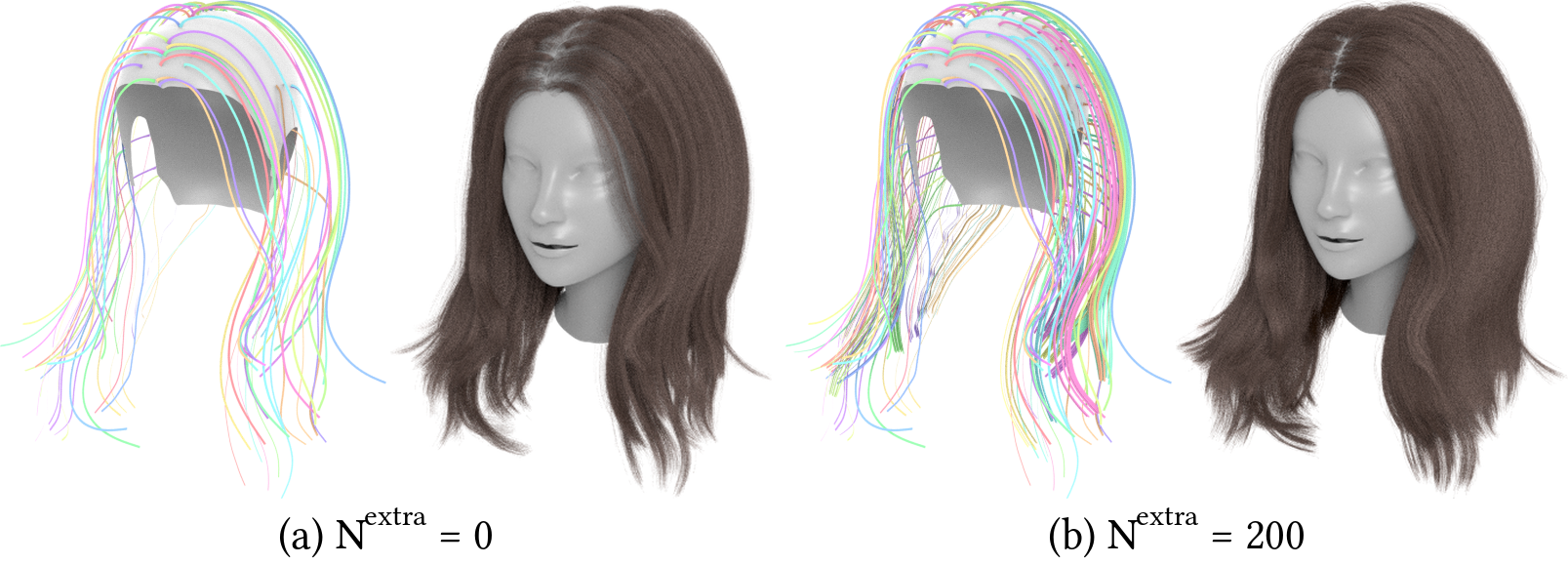}
    \caption{\textbf{Effect of the extra guides.} (a) Without extra guides ($N^\text{extra}=0$), only the card-seeded guides are used (left), and the strands are distributed unevenly, with visibly empty regions in the hair (right). (b) With our default budget ($N^\text{extra}=200$), the extra guides traced from FPS-sampled scalp roots (left) give smooth and uniformly distributed strands (right).}
    \Description{}
    \label{fig:extra_guides}
\end{figure}

\paragraph{Quantitative Measurement}
We quantitatively evaluate the ungroomed hairstyles using two types of metrics: a geometric metric that assesses the quality of the strand geometry itself, which degrades when strands gather into clusters or leave empty regions inside the hair volume, and image-space fidelity metrics that measure the structural similarity between the result and the input cards. First, we use the point-cloud Chamfer distance between the strands and the hair volume to assess how smoothly and evenly the strands are distributed within the hair volume. Since the strand vertices already form a point cloud, we uniformly sample an equal-size point set inside the hair volume and compute the Chamfer distance (denoted as CD) between the two point sets, with both sets capped at a common budget of 200K points so that the value is comparable across assets and across the ablation variants.  Second, we measure fidelity in image space on geometry AOVs rather than on shaded color. Cards and strands are shaded by different models, so a pixel comparison of their shaded renderings mostly measures this color misalignment instead of the geometric agreement. We instead render the per-pixel tangent and depth of both representations from the same camera (\autoref{fig:tangent_eval}). The tangent render encodes the local flow direction as a color, so it captures the wisp structure and the parting lines carried by the cards while being independent of hair color and lighting. We report PSNR and SSIM on both AOVs after cropping to the shared non-transparent bounding box and compositing over a constant background, together with the silhouette IoU of the two alpha masks as a measure of coverage. We do not use perceptual image metrics or a user study, because the AOV renders already remove the shading-model confound that such measures would have to control for. Detailed quantitative results are provided in \autoref{tab:evaluation}. Note that our method does not aim to exactly reproduce the appearance of the input cards, because our results tend to smooth out visual gaps between cards.
To complement the image metrics with a view-independent one, we also report the card-to-strand distance
{
\begin{equation}\label{eq:dcs}
    d_{c\to s} \;=\; \frac{1}{|\mathcal{P}_c|}\sum_{\pp\in\mathcal{P}_c}\;\min_{\qq\in\mathcal{S}^{\text{out}}}\|\pp-\qq\|,
\end{equation}}
where $\mathcal{P}_c$ contains points sampled uniformly on the opaque part of the card surfaces and $\qq$ runs over the output strand polylines. It is the mean distance from the cards to the closest strand, so it is independent of the sample count and card area and grows wherever strands drift away from or miss a card.
\paragraph{Ablation on Guides.}
Card-seeded guides cover only the scalp area beneath the root edges of the cards, and the extra guides of \autoref{sec:tracing} fill the rest. \autoref{tab:extra_guides} varies their budget $N^\text{extra}$ on the hairstyle of \autoref{fig:extra_guides}. We measure how evenly the guide roots cover the scalp by $\text{CoV}_\text{root} = \sigma(\delta) / \mu(\delta)$, where $\delta_j = \min_{k \neq j} d_{\mathcal{S}}(\rr_j, \rr_k)$ is the geodesic distance from root $\rr_j$ to its nearest other root along the scalp. Evenly spread roots give a low value, clustered roots a high one. Without extra guides, the wrappers leave part of the hair volume empty due to their limited resolution and expansion iterations. Each guide has to cover a larger region, which partitions both the scalp and the hair volume into fewer and larger wrapper cells. The initialized strands are then distributed unevenly in space, and both CD and $d_{c\to s}$ degrade (\autoref{tab:extra_guides}). The density relaxation cannot redistribute them uniformly and smoothly (\autoref{fig:extra_guides}), which indicates that well distributed guides remain necessary for our method. Beyond a small budget the two metrics no longer change, while $\text{CoV}_\text{root}$ keeps falling, so the roots keep spreading more evenly over the scalp. We use $N^\text{extra} = 200$ as the default.

\begin{table}[t]
  \centering
  \caption{\textbf{Extra-guide budget} on the hairstyle of \autoref{fig:extra_guides}. $N_\text{guide}$ is the guide count and $N^\text{extra}$ the extra-guide budget. CoV$_\text{root}$ is the coefficient of variation of the on-scalp geodesic nearest-neighbour distance between guide roots. CD and $d_{c\to s}$ are as in \autoref{tab:evaluation}, with CD measured against the hair volume of the default run.}
  \label{tab:extra_guides}
  \footnotesize
  \setlength{\tabcolsep}{4pt}
  {
  \begin{tabular}{rrrrr}
    \toprule
    $N^\text{extra}$ & $N_\text{guide}$ & CoV$_\text{root}\downarrow$ & CD$\downarrow$ & $d_{c\to s}\downarrow$ \\
    \midrule
    0   & 70  & 1.48 & 0.0093 & 0.00226 \\
    50  & 94  & 1.17 & 0.0079 & 0.00198 \\
    100 & 133 & 0.80 & 0.0079 & 0.00195 \\
    200 & 208 & 0.52 & 0.0078 & 0.00190 \\
    \bottomrule
  \end{tabular}}
\end{table}

\paragraph{Ablation Study on Full Strand Relaxation.}
We conduct a set of ablation experiments to demonstrate the effect of each term in the full-strand relaxation energy defined in Eq.~\eqref{eq:density_energy}. Without relaxation (\autoref{fig:density}~(c)), the initialized strands preserve the layered structure inherited from the cards, but exhibit overly clumped wisps with sharp, unnatural boundaries. This suggests that the initialization is reasonable, yet still requires further refinement. Removing the shape regularization term, as shown in \autoref{fig:density}~(d), causes strands to diffuse across wisp boundaries, thereby smoothing out the cluster structure that defines the hairstyle. Removing the boundary term (\autoref{fig:density}~(e)) allows strands to spread into empty space, making the overall silhouette difficult to preserve. With all terms enabled (\autoref{fig:density}~(f)), the Adam optimizer efficiently minimizes the loss and produces results that preserve both the transitions between wisps and the internal cluster structure. Our method continues to converge with additional iterations, and although more iterations may yield marginal improvements, our stopping criterion is sufficient to produce visually satisfactory results. \autoref{fig:loss} plots the relative loss over the iterations of the Adam optimizer.

\subsection{Comparison with Alternative Hair Synthesis Methods}
We compare our method with existing methods that synthesize full hair from guide hairs.

\begin{figure*}
    \centering
    \includegraphics[width=\linewidth]{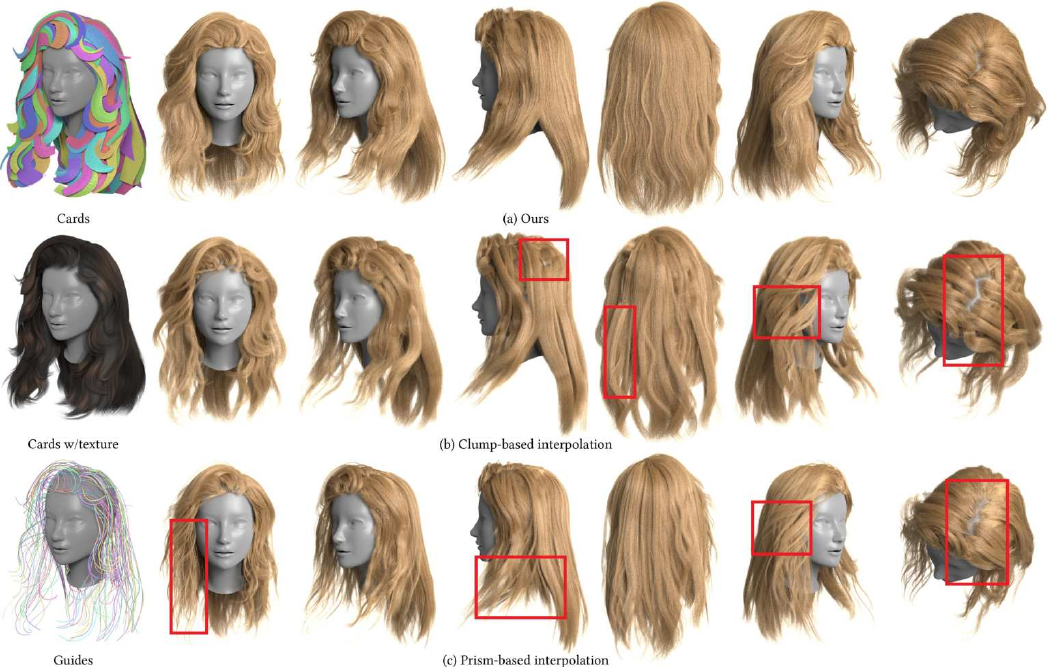}
    \vspace{-1.5em}
    \caption{\textbf{Comparison with Conventional Interpolation Methods.} (a) Our full hair synthesis method, (b) Clump-based/single guide interpolation, and (c) Prism-based/multi-guide interpolation.}
    \label{fig:interp}
    \Description{}
\end{figure*}

\paragraph{Comparison with Conventional Interpolation Methods.}
Given the guide hairs generated in~\autoref{sec:tracing}, two interpolation schemes are widely used in traditional industrial grooming systems, such as Maya XGen and Houdini, to generate full hairstyles from guide strands. These schemes are prism-based interpolation and clump-based interpolation. In prism-based interpolation, each dense strand root is interpolated from a triangular prism formed by neighboring guide strands. In clump-based interpolation, each strand is interpolated from a single neighboring guide using frame transport.
However, neither scheme produces usable results without labor-intensive post-editing. Prism-based interpolation blends across guides with conflicting features (\autoref{fig:interp}~(c)), which causes widespread mesh penetration and generates strands in regions where the hairstyle should instead contain gaps or follow a specific orientation. Clump-based interpolation preserves wispy features well, but often produces non-smooth and visibly clumpy results, with noticeable artifacts at clump boundaries (\autoref{fig:interp}~(b)).
In contrast, our method, shown in \autoref{fig:interp}~(a), combines the strengths of both approaches without requiring manual post-editing. Our wrapper-based interpolation and deformation preserve consistent features within each wrapper, while the relaxation step resolves clumpiness and discontinuities across wrapper boundaries. As a result, the generated strands are properly distributed under the boundary constraints, which treat each card as a thick wisp.
Both baselines are produced with the Hair Generate node of Houdini from the same guide set and hair count as ours, with parameters tuned by hand for the best result of each baseline and no manual editing of the strands afterwards; their settings are listed in Appendix~\ref{apdx:houdini}.

\begin{figure}
    \centering
    \includegraphics[width=1.0\linewidth]{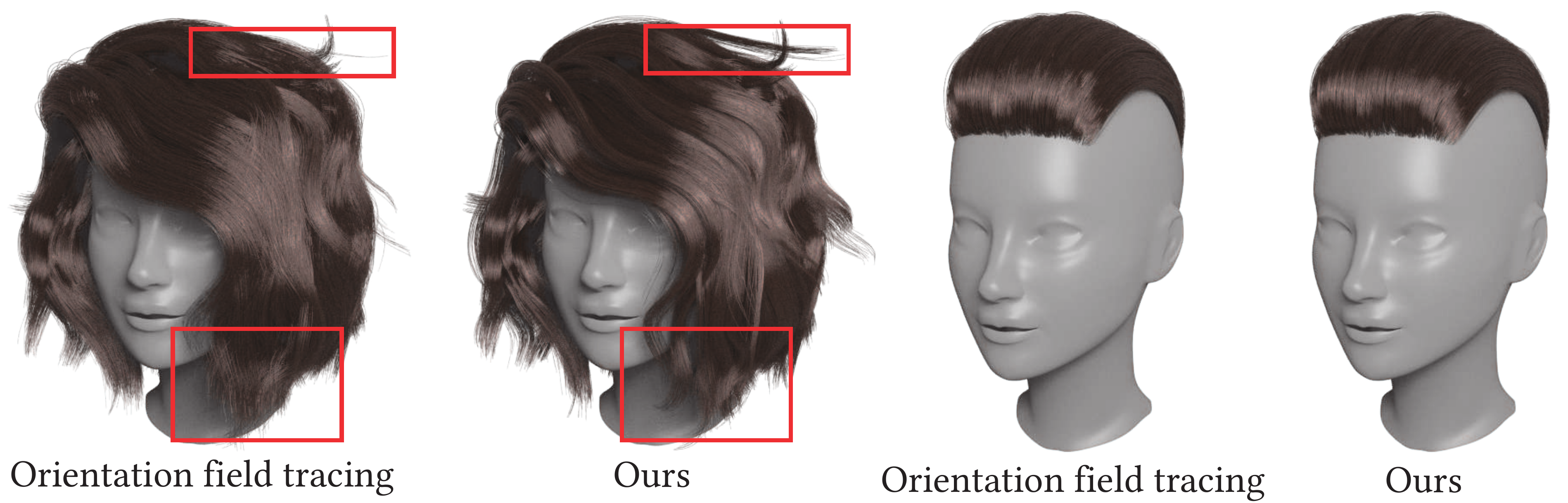}
    \caption{\textbf{Comparison on Orientation Field Tracing.} While both methods generate reasonable strands, our method better preserves the layered wisp structure, whereas orientation-field tracing tends to smooth out sharp features and is further limited by the grid resolution (the left two). For smooth, coherent hairstyles without flyaways, wisps, or bangs, the two methods yield similar results (the right two).}
    \Description{}
    \label{fig:orientation_field}
\end{figure}

\paragraph{Comparison with Orientation-Field Tracing.}
An alternative approach to hair synthesis is to trace strands along an orientation field, starting from roots on the scalp. This method naturally produces hairstyles that are well-connected to the scalp. To evaluate this alternative, we replace the full hair synthesis stage of our pipeline (\autoref{sec:full-hair}) with an orientation-field-based tracing scheme.
The 3D orientation field is constructed by rasterizing the guide strands and their interpolated strands within each individual wrapper, following~\cite{Wang09}, and then smoothly diffusing the field throughout the entire volume. As in~\cite{hu2015single}, strands are subsequently traced from roots uniformly sampled on the scalp. Although the resulting strands are smooth and free of overlap, this approach struggles to preserve the original clustering, layered structure, and per-card features (\autoref{fig:orientation_field}).
In addition, the original guides are not well-suited to serve as grooming operator guides in this setting, because traced strands may cross between clusters. Clumping and bending operators rely on each strand's relative position with respect to its center guide, and this assumption no longer holds when strands are allowed to move across clusters.

\begin{figure}[t]
    \centering
\setlength{\tabcolsep}{0pt}%
\begin{tabular}{@{}c@{\hspace{0.006\linewidth}}c@{\hspace{0.006\linewidth}}c@{\hspace{0.006\linewidth}}c@{}}
\includegraphics[width=0.2505\linewidth]{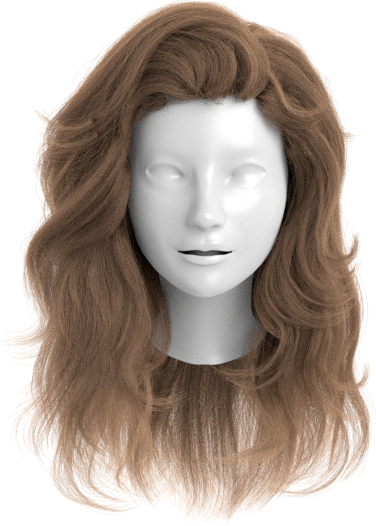} & \includegraphics[width=0.2492\linewidth]{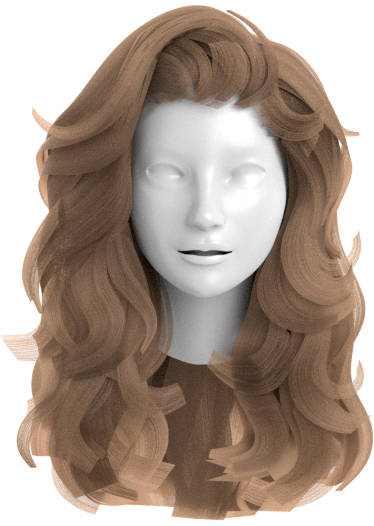} & \includegraphics[width=0.2474\linewidth]{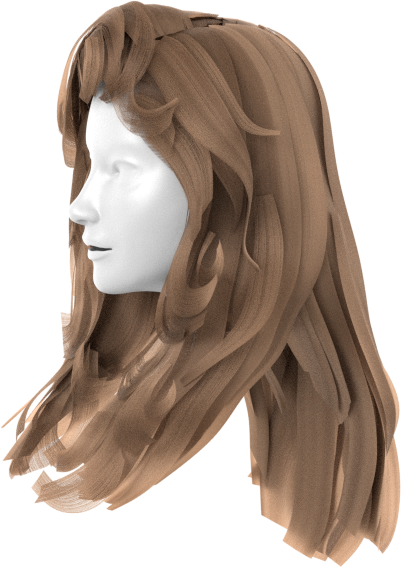} & \includegraphics[width=0.2350\linewidth]{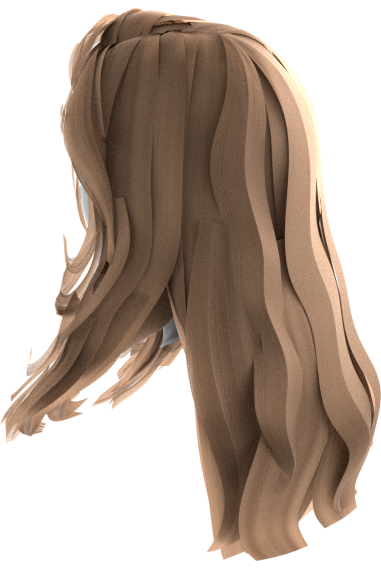}\\[2pt]
\scriptsize (a) Ours & \multicolumn{3}{c}{\scriptsize (b) Naive card tracing}\\[-2pt]
\scriptsize\textcolor{cdorange}{CD$\downarrow$: 0.0075} & \multicolumn{3}{c}{\scriptsize\textcolor{cdorange}{CD$\downarrow$: 0.0084}}
\end{tabular}

    \caption{\textbf{Naive card tracing against our result} on the same hairstyle as in \autoref{fig:interp}. (a) Our result fills the hair volume with scalp-rooted strands. (b) Strands traced directly on the cards: they lie on separate sheets with empty volume between them, and their roots sit on the card edges. \textcolor{cdorange}{CD$\downarrow$} is reported below each result.}
    \Description{}
    \label{fig:naive_tracing}
\end{figure}

\paragraph{Comparison with Naive Card Tracing.}
The most direct way to obtain strands from cards is to trace them on the textured strips. On every card we place strands at uniformly sampled lateral positions between its two hair-flow outlines, follow the outline family from the root edge to the tip, and lift the result to 3D through the card's UV mapping, with a strand budget proportional to the card area. This baseline is faithful to the cards by construction, and its card-to-strand distance is on par with ours. It inherits, however, the layout of the cards: the strands lie on separate sheets with empty volume between them, and their roots sit on the card edges rather than on the scalp (\autoref{fig:naive_tracing}). The Chamfer distance to a uniform sample of the hair volume exposes this. We reproduce this experiment on all examples of \autoref{tab:evaluation}: the same tracing gives CD between 0.0062 and 0.0132, against 0.0031 to 0.0063 for ours.

\begin{figure} [ht!]
    \centering
    \includegraphics[width=1.0\linewidth]{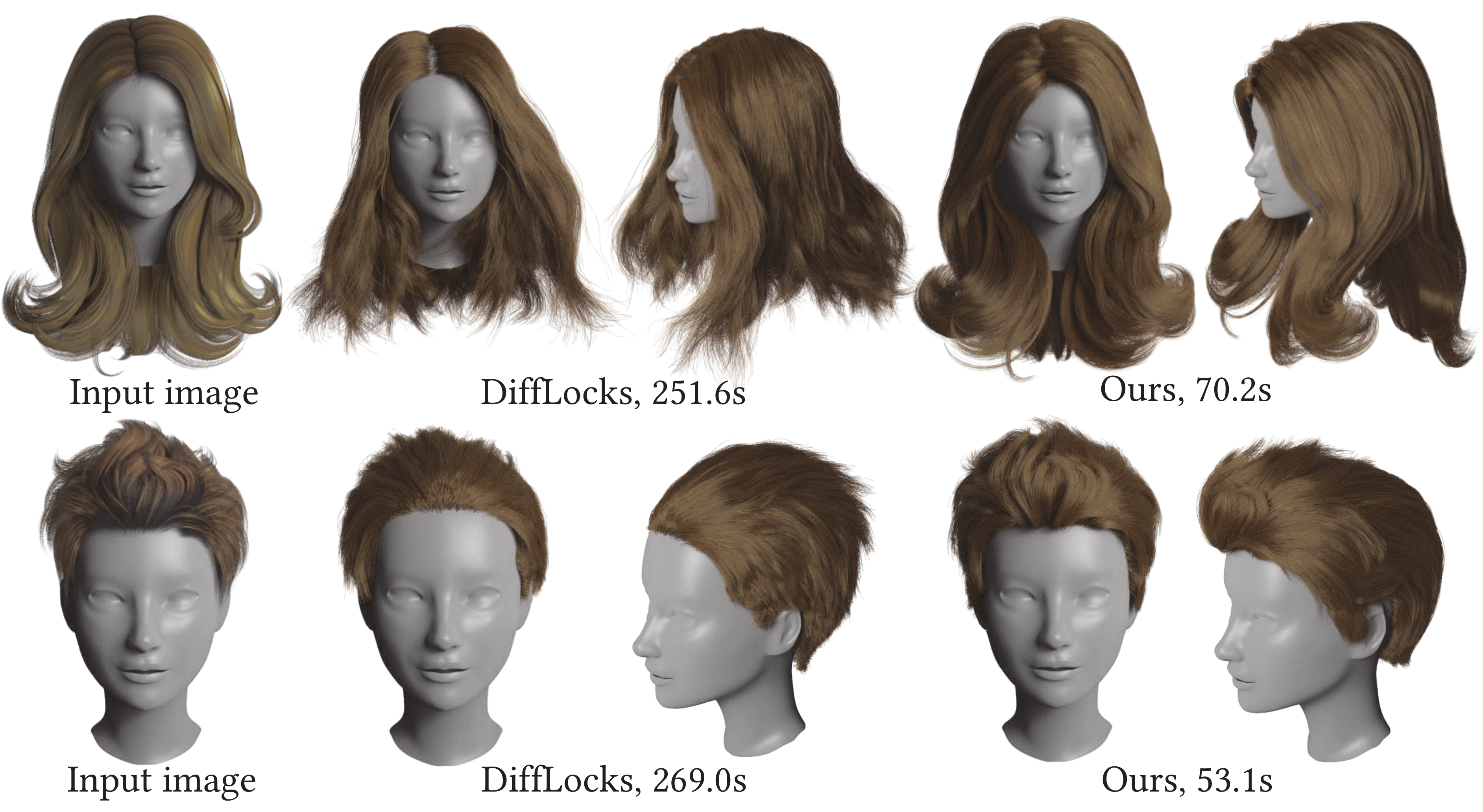}
    \caption{\textbf{Comparison with a Learning-based Image-to-strand Method.} While DiffLocks~\cite{difflocks2025} can generate reasonable results, our method is closer to the input, since we take richer geometric information into consideration.}
    \Description{}
    \label{fig:comp}
\end{figure}

\paragraph{Comparison with Learning-based Methods.}
Another end-to-end approach to hair strand generation is to directly synthesize strand models from an image using a neural network. Since the hair-card assets and textures are already available, one can render an image and use it as input to such a method. However, this approach faces two main challenges. First, when the bust is rendered with a stylized or representative avatar, methods that rely on facial landmarks may fail to reliably detect the face. Second, the rich geometric information present in the original asset is projected into image space, where much of it is inevitably lost.
While image-to-strand reconstruction is inherently challenging with limited input information, our pipeline complements it by fully exploiting the available geometry to recover the underlying structure of the hairstyle. We compare our method with DiffLocks~\mbox{\cite{difflocks2025}}. Although both methods produce reasonable results, our method remains closer to the input hairstyle by leveraging the richer geometric information available in the hair-card representation.

\subsection{Generalization to Production Assets}\label{sec:generalization}
\begin{figure*}[p]
    \centering
    \includegraphics[width=\textwidth]{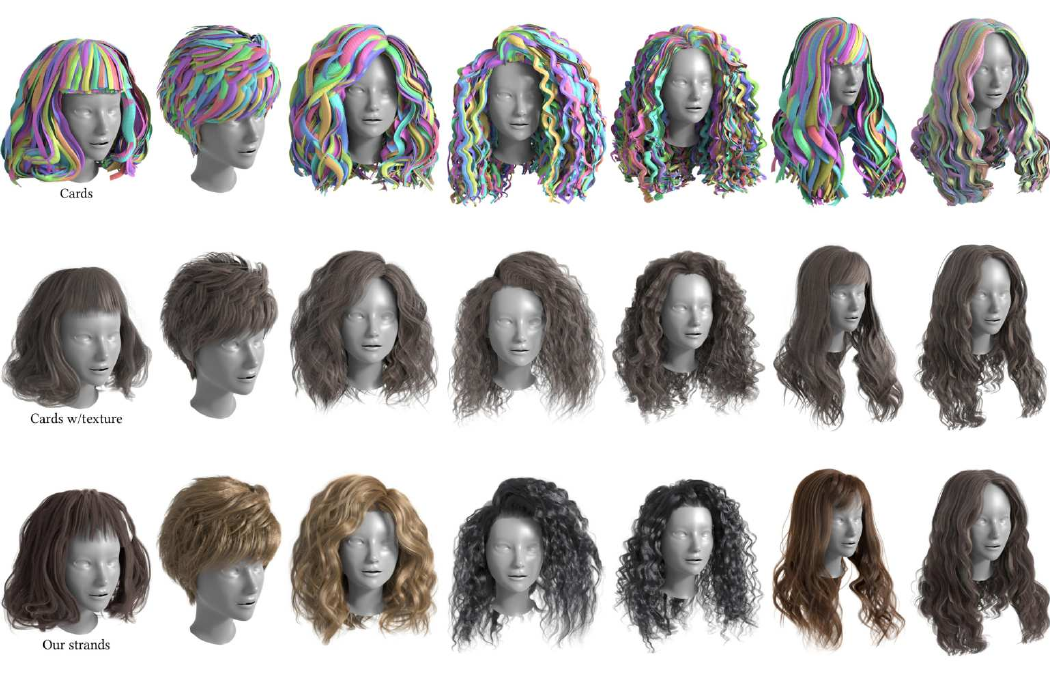}
    \vspace{-1.5em}
    \caption{\textbf{Cards in the wild.} We evaluate our pipeline on card-based models spanning diverse hairstyles from three additional in-the-wild asset packages. It produces satisfactory results even with hundreds to thousands of cards, and remains robust when the underlying card geometry is complex. Because most of these cards are strongly curved, we use 128 samples per strand for these assets.}
    \label{fig:card_in_the_wild}
    \Description{}
\end{figure*}

Beyond the curated dataset, we run the pipeline unchanged on artist-authored hair-card assets downloaded from three production sources, Character Creator 5, ArtStation, and the Unreal Engine asset store, all obtained under commercial licenses (\autoref{fig:card_in_the_wild}). These assets follow the industrial conventions for card layout and UV mapping, which differ from those of the dataset, and they contain hundreds to thousands of finely subdivided cards. None of this requires extra handling. The outline extraction of \autoref{sec:extraction} operates on each card's own UV region whether or not other cards share the same texture strip, duplicated and stacked cards collapse in the overlapping-guide reduction of the guide extraction stage, and every later stage is untouched. The only parameter we change is the number of samples per guide and per strand, which we raise to 128 because these cards are strongly curved along their length. All other parameters are those of \autoref{tab:hparams}. Processing takes 6 to 22 minutes per asset on one RTX~4090. As \autoref{fig:card_in_the_wild} shows, the reconstructed strands match the geometry that the cards provide.

\subsection{Texture-Guided Initialization}\label{sec:texture_guided}
\begin{figure*}[p]
    \centering
    \begin{minipage}[b]{0.280\textwidth}
        \centering
        \includegraphics[width=\linewidth]{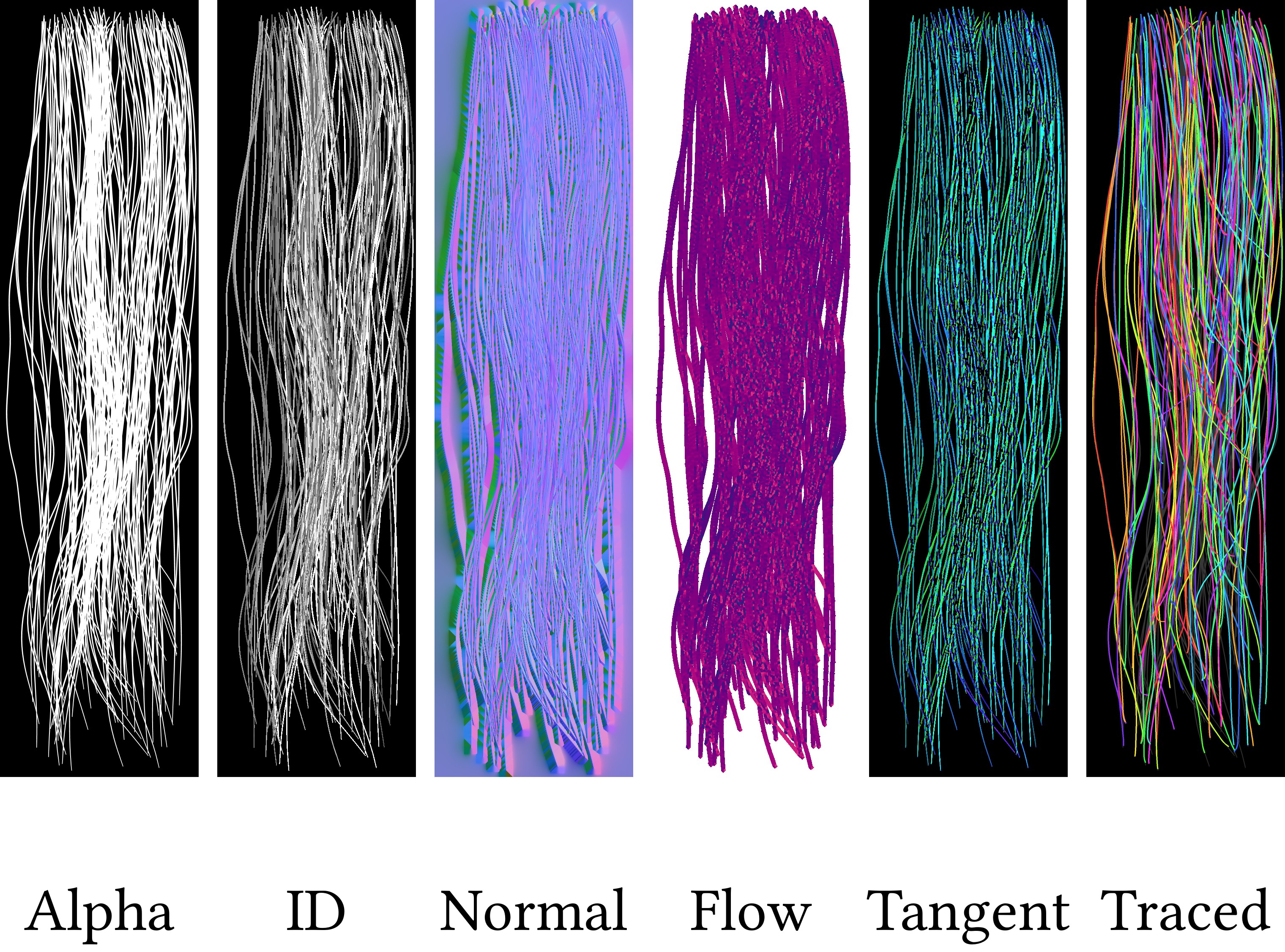}\\[2pt]
        {\footnotesize (a)}
    \end{minipage}\hfill
    \begin{minipage}[b]{0.710\textwidth}
        \centering
        \includegraphics[width=\linewidth]{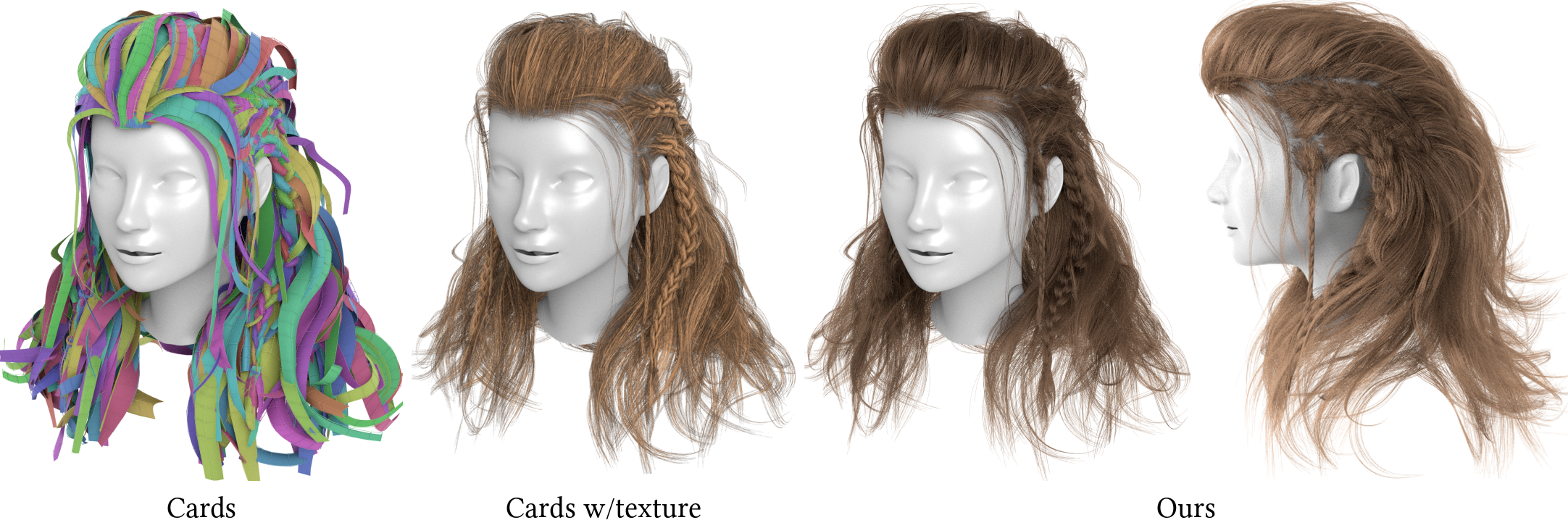}\\[2pt]
        {\footnotesize (b)}
    \end{minipage}
    \caption{\textbf{Texture-guided initialization.} (a) One texture strip of the test asset, from left to right: its alpha mask, its hair-ID map, its normal map, its flow map, its tangent map (hue encodes the direction), and the strands traced along the tangent map, one color per strand. (b) The input cards, the cards with their textures, and our strand result. When an asset ships high-fidelity textures such as an ID map and a flow map, we use their tangent information to reconstruct strands that follow the on-card structure and density more closely. The result closely matches the rendered textured cards.}
    \Description{}
    \label{fig:texture_guided}
\end{figure*}

Some high-quality assets ship a flow map with their cards, a texture that stores the hair direction at every pixel of a strip (\autoref{fig:texture_guided}~(a)). The guide extraction of \autoref{sec:extraction} reads the texture only to locate the root and tip of a strip. When an asset provides this information, we can run a tracing stage that uses it. A tracer combines the ID map and the flow map into a tangent map, grows streamlines along it inside the alpha mask, connects them into long strands, smooths them, and lifts them to 3D through the UV mapping of every card that uses the strip. The test asset is an artist-authored hairstyle with 1,099 cards and 13 strips, fitted to our bust.

These strands are not a hair model: each lies on its card, so the set is a stack of sheets with the same voids as the cards, and every card that shares a strip receives the same copy. Taking all of them as guides would also give far too many guides. We therefore use the traced strands in the binding stage and keep one traced guide per card, the medoid of the card's strands, for the wrapper stage, so the number of wrappers does not grow. Inside each wrapper, the traced strands serve as the initial state of the strands and as the density target of the relaxation. \autoref{fig:texture_guided}~(a) shows the stages on one strip: the ID map and the flow map give the tangent map, and the strands are traced on the 2D strip along it. \autoref{fig:texture_guided}~(b) shows that the result closely matches the rendered textured cards. The price is time. Generating the tangent map and tracing the strips takes about 12 minutes on this asset and binding the traced strands to the scalp about 4 minutes, so the run on this asset takes about three times the 8 minutes of our original pipeline on the same case.

We do not explore this route or its acceleration further, since most assets in our dataset ship only an RGB color texture and no high-fidelity maps. We make no assumption about the textures an asset provides, and leave high-fidelity reconstruction that matches the rendered result to future work.

\section{Applications}

This section demonstrates three applications from the output of our method.

\newpage  
\subsection{Hairstyle Blending}\label{sec:blending}
Because strand-based hair models are expensive and time-consuming to obtain, a common strategy in neural hair reconstruction is to enlarge the dataset by blending existing hairstyles.

The strand-based hairstyles produced by our pipeline are traceable via guide strands and are scalp-rooted by construction, making them composable in a way that the original hair cards are not. We exploit this property with a template-blend operator that merges two converted assets into a single coherent hairstyle by replacing selected scalp regions between them, e.g., ``replace the candidate's bangs with those of the target hairstyle.'' Such an operator is natural for strand representations but infeasible for card-based assets, because cards encode only the overall hair geometry and do not explicitly represent where each strand belongs on the scalp.

\begin{figure} [ht!]
    \centering
    \includegraphics[width=1.0\linewidth]{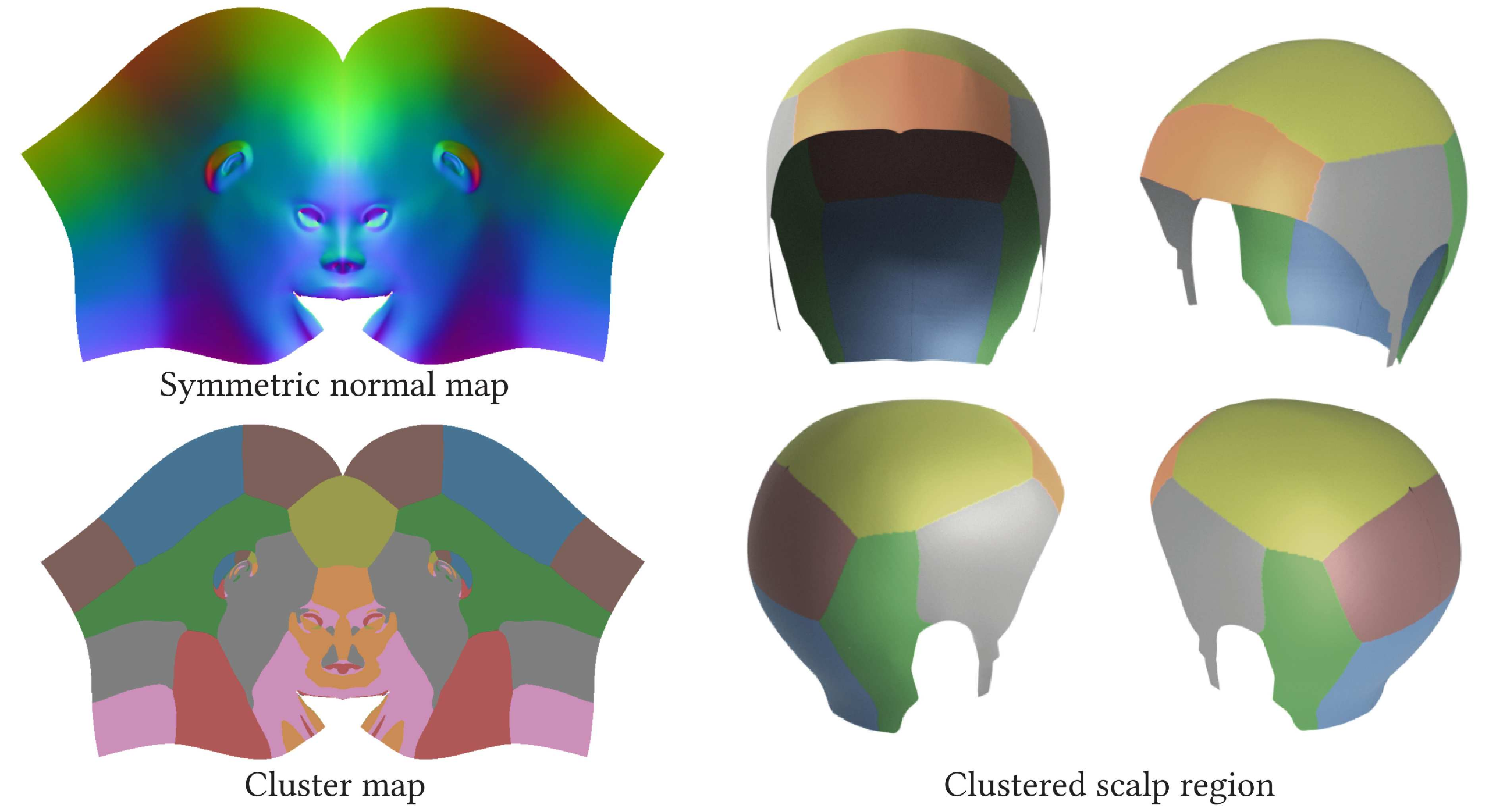}
    \caption{\textbf{Scalp Region Clustering and Semantic Labeling.} We first perform bilateral clustering over the scalp region and then assign semantic labels based on the clustering result. The left image shows the symmetric normal map used for K-means clustering, which produces a symmetric partition of the scalp. We then use this partition to assign semantic labels to the scalp regions.}
    \Description{}
    \label{fig:scalp_cluster}
\end{figure}

To enable blending, we first run both input assets through the full conversion pipeline, including guide extraction, binding, and tracing. This produces two scalp-connected guide sets defined on the same scalp $\mathcal{S}$ and bust $\mathcal{B}$. We then partition $\mathcal{S}$ into eight semantic regions---frontal, left/right frontal, parietal, left/right temporal, vertex, and occipital---using K-means clustering on smoothed scalp normals. For bilaterally symmetric regions, we further split each cluster into left and right halves using face-graph connected components, so that non-median regions receive distinct labels. Each guide is assigned to a semantic region according to the cluster label at its scalp root (see \autoref{fig:scalp_cluster}).

The user specifies replacement rules of the form $\rr_{\text{target}} \leftarrow \rr_{\text{candidate}}$. For each rule, we remove the target guides in $\rr_{\text{target}}$ and replace them with the candidate guides from $\rr_{\text{candidate}}$. To maintain a clean transition across the seam, we apply an orientation-alignment safeguard. Specifically, an orientation-compatibility filter removes any imported guide whose root tangent is inconsistent with those of its $k$ nearest target-side neighbors on the scalp. In our implementation, we use $k = 8$ and a cosine similarity threshold of $\tau_{\text{ori}} = 0.7$. This prevents abrupt tangent discontinuities at region boundaries.

\begin{figure}
    \centering
    \includegraphics[width=1.0\linewidth]{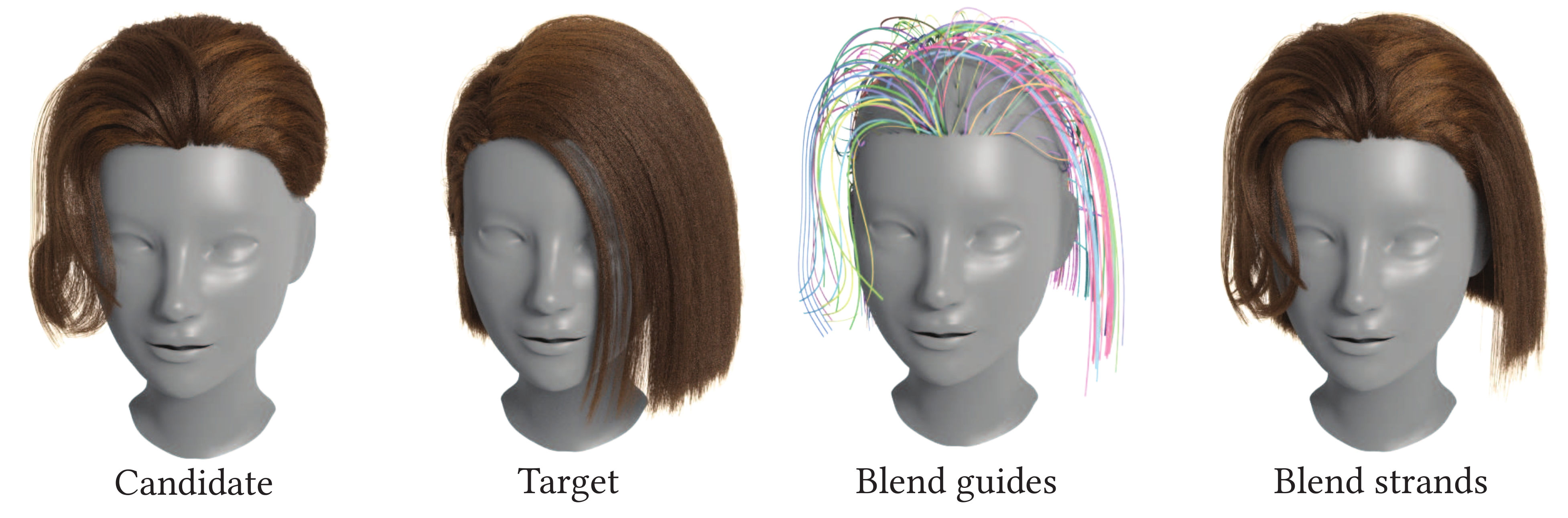}
    \caption{\textbf{Examples on Template Blending.} We expand our database by blending hairstyles in a traceable, semantically meaningful manner, with an orientation-compatibility safeguard. The example above illustrates the replacement rule along with the candidate and target hairstyles. In this case, the replacement rule replaces the target's frontal and left frontal regions with those of the candidate. }
    \label{fig:template_blend}
\end{figure}

After guide-level template blending, all downstream stages remain unchanged, including card-mesh merging based on guide assignment, wrapper construction, relaxation, and dense strand synthesis. As long as the replacement has a meaningful semantic interpretation, this method robustly blends hairstyles while preserving spatial coherence, thereby providing an effective way to enlarge the hairstyle database. \autoref{fig:template_blend} shows several blending results.

\subsection{Secondary Editing via Groom Modifiers}

Our scalp-connected strand representation is compatible with conventional grooming pipelines, such as Maya XGen and Houdini. In particular, compatibility with Houdini's hair-grooming system can be achieved by simply computing per-strand barycentric coordinates on the scalp surface. In addition, we implement bend, scale, and fuzz modifiers, as described in detail in~\cite{Chang2025}.

\begin{figure}[ht!]
    \centering
    \includegraphics[width=\linewidth]{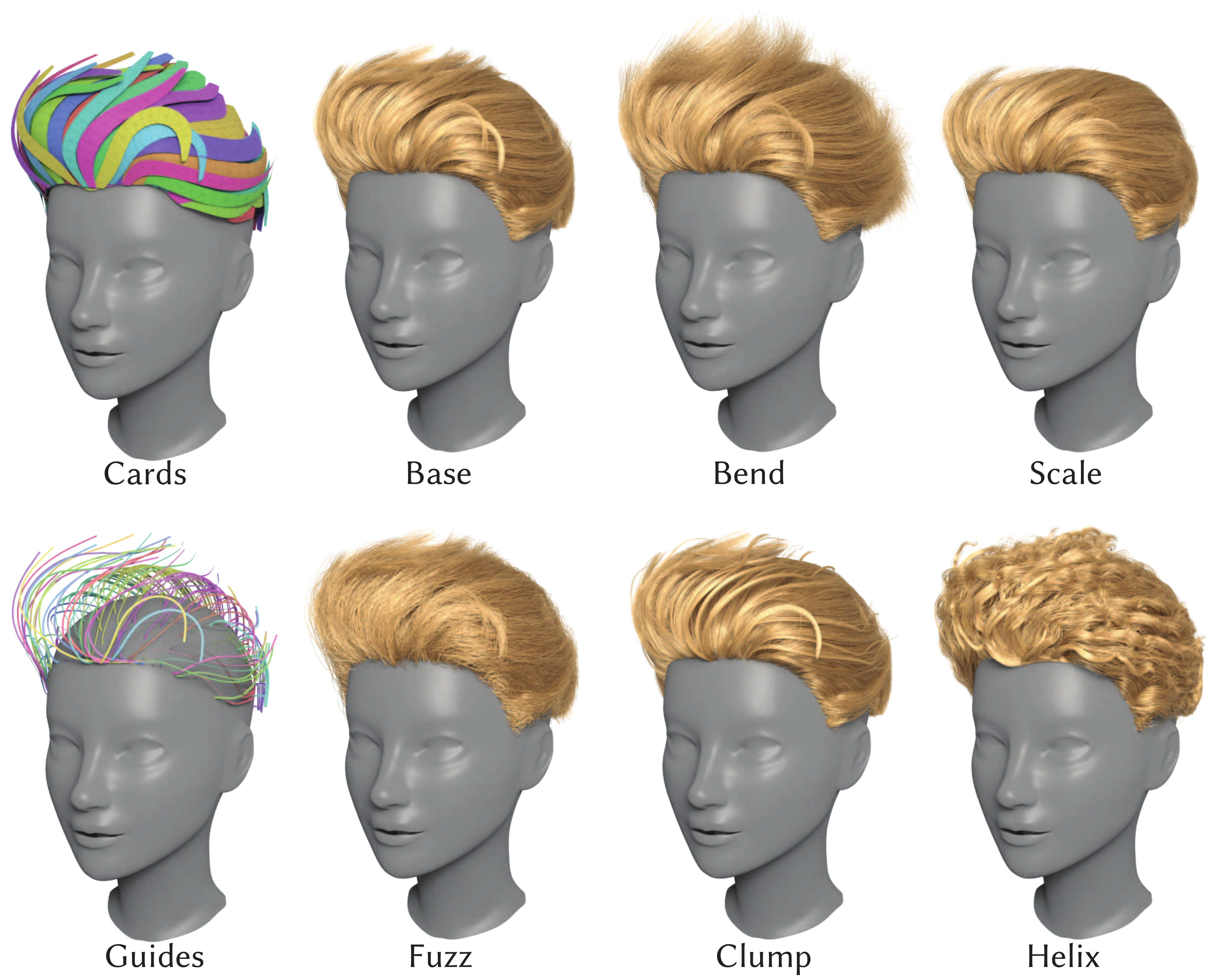}
    \caption{\textbf{Secondary Hair Editing.} Our pipeline is compatible with secondary grooming edits. The guides within the wrapper can serve as guides for the grooming operators. In addition, for the curly/helix operator, we detail the process in \autoref{apdx:wrapper_deformation} and \autoref{apdx:helix_operator}. Hair grooming offers yet another way to expand the hair strand database with richer features.}
    \Description{}
    \label{fig:groom}
\end{figure}

\begin{figure}[ht!]
    \centering
    \includegraphics[width=\linewidth]{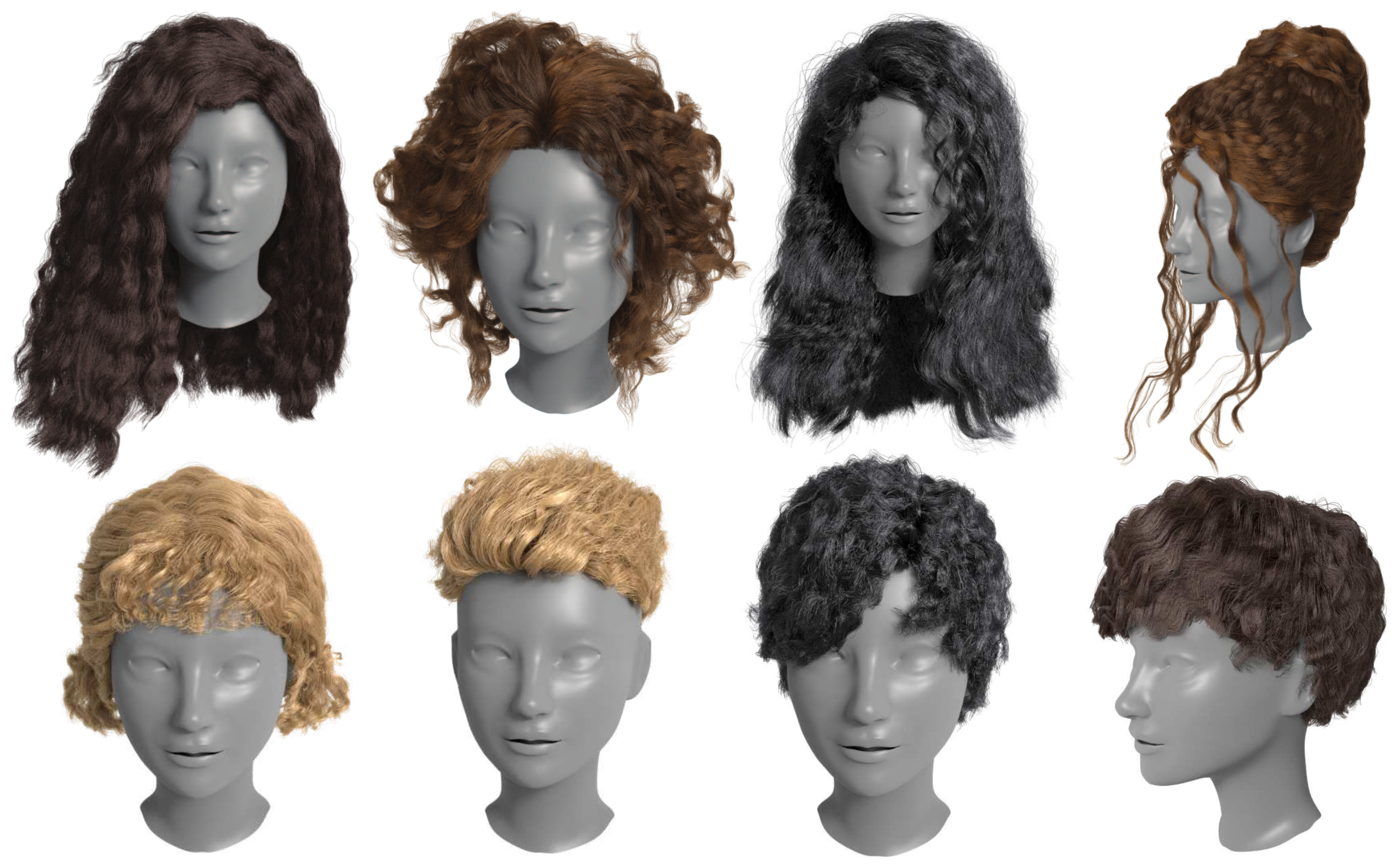}
    \caption{\textbf{Curly/helix Modifier.} We design a curly/helix operator compatible with our pipeline that enhances the base hair features.}
    \Description{}
    \label{fig:curlys}
\end{figure}

For curly and helical hairstyles, we further extend the system by deforming the guide strands and generating the base hair through rotational transport, followed by the wrapper deformation procedure described in \autoref{sec:full-hair} and \autoref{apdx:wrapper_deformation}. \autoref{fig:groom} presents several examples of grooming applied to the base hair generated by our method. Because the wrapper captures the volumetric distribution of each wisp, these modifiers can generate a wide variety of hairstyles while preserving the characteristic shape features of the original base hair.
For additional details on the curly and helical grooming modifiers with wrapper deformation, please refer to \autoref{apdx:helix_operator}. \autoref{fig:curlys} shows further results of hairstyles groomed using these modifiers.

\subsection{Strand-based Hair Simulation and Rendering}
\begin{figure}[ht!]
    \centering
    \includegraphics[width=\linewidth]{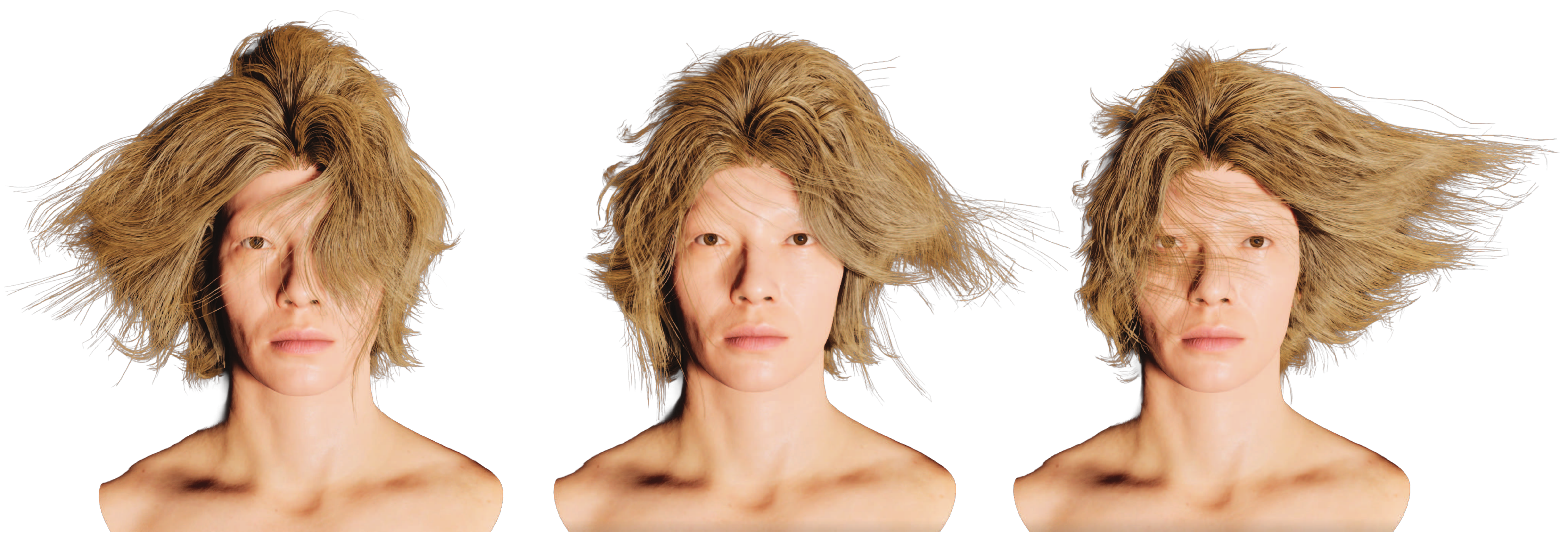}
    \caption{\textbf{Simulation and Rendering in Unreal Engine~\cite{unrealengine}.} We import the strand-based output generated by our pipeline into the game engine and drive it with a periodic wind force. The engine handles the scenario robustly and produces convincing simulation results.}
    \Description{}
    \label{fig:UE}
\end{figure}

Because our output guarantees scalp-connected strands, it can be fed directly into a strand-based simulator without any additional processing. \autoref{fig:UE} shows an example of our result simulated and rendered in Unreal Engine~\cite{unrealengine} in real time, where we apply a periodic wind force as external excitation. The engine handles the scenario robustly and produces convincing results.

\section{Conclusion and Limitations}
We proposed a five-step pipeline for converting hair cards into high-fidelity hair strands. Our method addresses two key challenges: the spatial occupancy when lifting ribbon-like cards into volumetric wisps, and the layered structure that emerges when treating a hairstyle as a whole. We developed a wrapper-based relaxation scheme that relaxes the guide strands while simultaneously expanding the volume of each wisp hosted by them.  The full hair is then synthesized within the wrapper mesh and refined through a subsequent relaxation pass. Our strand-based hair is further groomable with a variety of modifiers, which enrich the hairstyle with features such as curls, waves, and helical structures. Combined with our hairstyle blending method, this pipeline enables us to construct a dataset of $50$K diverse hairstyles. 
This complete pipeline thus bridges the gap between low-cost hair card assets and high-fidelity strand-based grooms, enabling the conversion of existing card-based content into rich strand-based models suitable for simulation, learning-based hair modeling datasets, and data-driven applications.

\paragraph{Limitations.} While our method focuses on the volumetric representation and the inherent layered structure of a hairstyle, it does not account for the strand distribution within a wisp. This includes the strand density and the cross-sectional variation along the centerline. Reconstructing strands from card textures remains another challenging problem in its own right. We further show that an asset shipping high-fidelity textures with tangent information can be reconstructed to match its rendering more closely. Optimizing and accelerating that stage, and extending it to RGB-only textures, are left to future work. In addition, our method does not guarantee a correct topology for braids, as it extracts guides directly from the cards. For assets that blend volumetric hair meshes with cards, our method cannot deliver the quality we otherwise promise, since it relies on the 2D parameterization of the card geometry to extract the correct centerline. Reconstructing the strands for such hybrid card-and-mesh assets is a natural next step.

\bibliographystyle{ACM-Reference-Format}
\bibliography{ref}
\appendix
\section{Wrapper Relaxation Algorithm}\label{apdx:relaxation}

In this section, we provide additional details on the wrapper relaxation algorithm presented in~\autoref{alg:tube_relaxation}. The wrapper set $\Omega=\bigcup_{j=1}^{|\mathcal{G}|}\mathcal{W}_j$ is an approximate partition of the hair volume generated by the expansion algorithm.

\begin{algorithm}[!htbp]
\caption{Wrapper-constrained relaxation.}
\label{alg:tube_relaxation}
\KwIn{$\Phi_\Omega,\ \Phi_\mathcal{B},\ \mathcal{S},\ \mathcal{R},\ \Gamma,\ \mathcal{G},\ \Pi:\mathcal{G}\!\to\!\mathcal{R}$ }
\KwOut{$\mathcal{G}^\text{relax},\ \{\mathcal{W}_j\}$ \tcp*[r]{relaxed guides, per-guide wrappers}}
\For(\tcp*[h]{wrapper initialization}){$j = 1, \dots, |\mathcal{G}|$}{
    $P_j \leftarrow$ Voronoi cell of $\Pi(\GG_j)$ on $\mathcal{S}$, clipped by $\Gamma$\;
    $\mathcal{W}_j \leftarrow \textsc{BuildPerGuideWrapper}(P_j,\, \GG_j)$\;
}
\For{$k = 1, \dots, K_\text{relax}$}{
    \ForEach(\tcp*[h]{outward cross-sectional normal at $v$}){ring vertex $v$ in any cross-section}{
        $\cc_v \leftarrow$ centerline point on $\GG_j$ at the layer of $v$\;
        $\nn_v \leftarrow (\vvv - \cc_v) \,/\, \|\vvv - \cc_v\|$, $\;s_v \leftarrow \Delta r$ \tcp*{normal expansion}
    }
    \ForEach{ring edge $(u, w)$}{
        $\uu\!\to\!\uu+s_u\nn_u$, $\;\ww\!\to\!\ww+s_w\nn_w$\\
        \tcp*[h]{ACCD computes the time of impact: $t^{\text{TOI}}_{uw} \in [0,1]$}\\
        $t^{\text{TOI}}_{uw} \leftarrow \textsc{ACCD}\bigl(\uu, \ww,\, s_u\nn_u,\, s_w\nn_w;\,\Phi_\Omega, \Phi_\mathcal{B}\bigr)$\;
        $s_u \leftarrow \min(s_u, t^{\text{TOI}}_{uw}\,\Delta r)$, \quad $s_w \leftarrow \min(s_w, t^{\text{TOI}}_{uw}\,\Delta r)$\;
    }
    $s^\text{ceil}_v \leftarrow s_v$ for every ring vertex \tcp*{ACCD ceiling for smoothing}
    \For(\tcp*[h]{$K_\text{smooth}$ Laplacian sweeps along the tube longitudinal direction}){$i = 1, \dots, K_\text{smooth}$}{
        \ForEach{ring vertex $v$}{
            $\bar{s}_v \leftarrow \frac{1}{|\mathcal{N}(v)|}\sum_{u \in \mathcal{N}(v)} s_u$ \tcp*{$\mathcal{N}(v)$ = one-ring longitudinal neighbors along the tube}
            $s_v \leftarrow \min\bigl((1 - \alpha)\,s_v + \alpha\,\bar{s}_v,\;\; s^\text{ceil}_v\bigr)$\;
        }
    }
    \ForEach(\tcp*[h]{advance every vertex}){ring vertex $v$}{
        $\vvv \leftarrow \vvv + s_v\,\nn_v$\;
    }
    \ForEach{$\GG_j \in \mathcal{G}$}{
        \lIf{$|\GG_j| < l_\text{spline}$}{$\GG_j \leftarrow \textsc{CubicSplineFit}(\GG_j)$}
        \tcp{collect blocked displacement}
        \For(\tcp*[h]{each cross-sectional ring}){$k = 1, \dots, |\GG_j| - 1$}{
            $\dd_{j,k} \leftarrow \tfrac{1}{T}\!\sum_{t=1}^{T}\!\bigl(s_{v^t_{j,k}} - \Delta r\bigr)\,\nn_{v^t_{j,k}}$
            \tcp*{negated mean blocked displacement over the $T$ ring samples}
        }
        $\hat{\GG}_j \leftarrow \GG_j + \dd_j$\tcp*{predicted position}
        $\GG_j \leftarrow \textsc{XPBDRod}\bigl(\GG_j;\, \hat{\GG}_j\bigr)$\tcp*{rod constraints}
        $\GG_j \leftarrow \textsc{ACCDUpdate}(\GG_j;\, \Phi_\Omega,\, \Phi_\mathcal{B})$\;
    }
    \ForEach{$\mathcal{W}_j$}{
        $\mathcal{W}_j \leftarrow \textsc{ResyncRings}(\mathcal{W}_j,\, \GG_j)$ \tcp*{wrapper--guide re-synchronization by guide movement}
        $\mathcal{W}_j \leftarrow \textsc{ACCDUpdate}(\mathcal{W}_j;\, \Phi_\Omega,\, \Phi_\mathcal{B})$\;
    }
}
\Return $\mathcal{G}^\text{relax} = \{\GG_j\}, \{\mathcal{W}_j\}$\;
\end{algorithm}

\paragraph{XPBD rod solver.}

\begin{figure}[t]
    \centering
    \includegraphics[width=0.8\linewidth]{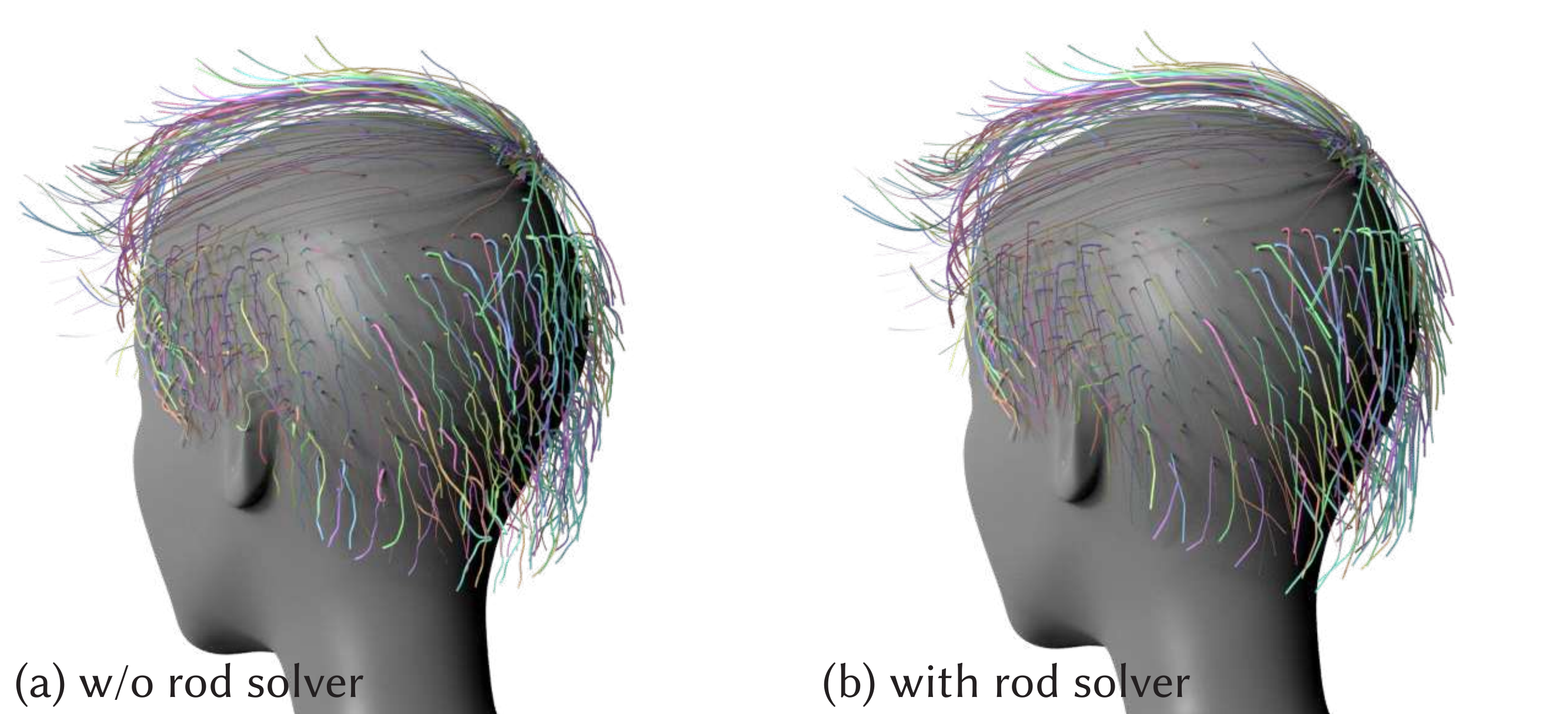}
    \caption{\textbf{Comparison with and without the rod solver.} (a) Vertex-wise refinement. (b) Refinement with the XPBD rod solver.}
    \Description{}
    \label{fig:rod_vs_vertex}
\end{figure}

The guide strands are relaxed under collisions imposed by the wrapper meshes. Because the wrapper meshes are expanded iteratively and clamped by ACCD, collision detection provides a blocked displacement for each cross-section of each wrapper. We use this blocked displacement to drive the guide dynamics by updating the predicted positions of the guide strands accordingly. We then model the guide strands as elastic rods and solve them with an XPBD solver. This avoids the noise and implausible deformations that arise when refining guide strands independently on a per-vertex basis (see~\autoref{fig:rod_vs_vertex}).

We represent strand $\GG_j$ by its vertices, $\GG_j=(\gg_0,\cdots,\gg_K)$, and its segment angles, $\Theta_j=(\theta_0,\cdots,\theta_{K-1})$. The optimization energy for each strand is defined as
\begin{equation}
    E^\text{rod}=\frac{M}{h^2}||\GG_j-\hat{\GG_j}||^2 + w^\text{len}||\mathrm{C}^\text{len}(\GG_j)||_F^2 + w^\text{ang}||\mathrm{C}^\text{ang}(\Theta_j)||_F^2,
\end{equation}
where the first term is the momentum term, followed by the segment-length constraint vector and the rest-angle constraint vector. Specifically,
$\mathrm{C}^{\text{len}}_k$$=||\gg_{k+1}-\gg_{k}||-l^\text{rest}_k$ and $\mathrm{C}^{\text{ang}}_k=\theta_{k}-\theta_{k}^\text{rest}$, where $(\bullet)^\text{rest}$ denotes the corresponding rest-shape measurements. The stiffness values and the number of sweeps are listed in \autoref{tab:hparams}. All strands are processed independently and in parallel.

\section{Wrapper-based Hair Deformation}\label{apdx:wrapper_deformation}

Here we formalize the wrapper deformation used during dense strand synthesis in \autoref{sec:full-hair}. After wrapper relaxation, each guide $j$ is associated with a smooth centerline $\GG_j$, a volumetric tube-shaped wrapper $\mathcal{W}_j\subset\Omega$ with boundary $\partial\mathcal{W}_j$, and a scalp region $\mathcal{R}_j \subset \mathcal{S}$ containing the root of the guide, $\Pi(\GG_j)$. Full-hair synthesis converts each such triplet into a dense scalp-connected wisp whose strands originate from the scalp, fill the wrapper uniformly, and remain entirely inside $\mathcal{W}_j$.

A naive strategy would be to translate each scalp root along $\GG_j$. However, this generally violates the containment condition when the wrapper cross-section is non-circular or varies along the guide. Instead, we formulate synthesis as a per-cross-section deformation from a canonical unit disk to the actual wrapper cross-section. The process consists of two stages: first, parallel transport of canonical scalp coordinates along the guide; second, radial fitting of the disk to the wrapper boundary at each cross-section.

\paragraph{Problem statement.}
Let $\GG_j(s): [0, L_j] \to \mathbb{R}^3$ be the arc-length parameterization of guide $j$, where $\GG_j(0) \in \mathcal{S}$, and let $\{\TT_j(s), \NN_j(s), \BB_j(s)\}$ denote its parallel-transport frame anchored at $s = 0$. We write $\partial\mathcal{W}_j$ for the closed boundary surface of the per-guide wrapper, and $\mathcal{P}_j(s)$ for the cross-sectional plane passing through $\GG_j(s)$ with normal $\TT_j(s)$.

Given the triplet $(\GG_j,\, \partial\mathcal{W}_j,\, \mathcal{R}_j)$, our goal is to produce a dense strand set $\{\SS_i\}$ such that each strand is rooted in $\mathcal{R}_j$ and remains inside $\partial\mathcal{W}_j$.

\paragraph{Cross-sectional radius field.}
The boundary $\partial\mathcal{W}_j$ defines a scalar radius field $R_j: [0, L_j] \times [0, 2\pi) \to \mathbb{R}^+$, where $R_j(s, \theta)$ denotes the distance from $\GG_j(s)$ to $\partial\mathcal{W}_j$ along the in-plane unit ray
\begin{equation}\label{eq:wd_ray}
    \ee_j(s, \theta) = \cos\theta\,\NN_j(s) + \sin\theta\,\BB_j(s).
\end{equation}

\begin{wrapfigure}[10]{r}{0.4\linewidth}
  \centering
  \resizebox{\linewidth}{!}{
\begin{tikzpicture}[black, yscale=0.78, every node/.style={font=\scriptsize, inner sep=1pt}, >=stealth, line cap=round, line join=round]
  \definecolor{slab}{RGB}{0,128,0}
  \definecolor{etac}{RGB}{255,120,0}
  \tikzset{blob/.style={fill=gray!12, draw=gray!70, line width=0.5pt}}
  \draw[blob] plot[smooth cycle, tension=0.8] coordinates {(0.1,0.55) (0.55,0.9) (1.3,0.75) (2.15,0.9) (2.95,0.55) (2.6,0.12) (1.5,0.02) (0.65,0.08)};
  \draw[blob] plot[smooth cycle, tension=0.8] coordinates {(0.3,2.0) (0.75,2.32) (1.45,2.18) (2.2,2.34) (2.8,2.0) (2.5,1.66) (1.5,1.57) (0.75,1.61)};
  \draw[blob] plot[smooth cycle, tension=0.8] coordinates {(0.6,3.5) (1.0,3.78) (1.7,3.66) (2.3,3.78) (2.75,3.5) (2.45,3.2) (1.7,3.13) (1.05,3.2)};
  \draw[slab, dashed, line width=0.5pt] plot[smooth, tension=0.6] coordinates {(0.1,0.55) (0.3,2.0) (0.6,3.5)};
  \draw[slab, dashed, line width=0.5pt] plot[smooth, tension=0.6] coordinates {(2.95,0.55) (2.8,2.0) (2.75,3.5)};
  \draw[line width=0.9pt, ->] (1.3,0.4) .. controls (1.35,1.0) and (1.35,1.5) .. (1.4,1.95)
        .. controls (1.5,2.5) and (1.65,3.0) .. (1.75,3.5) .. controls (1.8,3.7) and (1.85,3.8) .. (1.92,4.0);
  \fill (1.3,0.4) circle (1.3pt);
  \fill (1.75,3.5) circle (1.3pt);
  \coordinate (G) at (1.4,1.95);
  \coordinate (X) at (2.3,2.22);
  \draw[gray!70, dashed, line width=0.5pt] (G) -- (X);
  \draw[blue, ->, line width=0.9pt] (G) -- ($(G)!0.42!(X)$);
  \node[blue, anchor=north west, inner sep=0pt] at ($(G)+(0.12,-0.03)$) {$\ee_j(s,\theta)$};
  \fill (G) circle (1.4pt);
  \node[anchor=east] at ($(G)+(-0.05,0.08)$) {$\GG_j(s)$};
  \fill[red] (X) circle (1.4pt);
  \node[anchor=north] at ($(X)+(0.02,-0.04)$) {$\xx$};
  \node[etac, anchor=south west, inner sep=0pt] at (1.74,2.36) {$\eta R_j(s,\theta)$};
  \node[slab, anchor=west] at (3.08,3.5) {$s=L_j$};
  \node[slab, anchor=west] at (3.08,1.95) {$s$};
  \node[slab, anchor=west] at (3.08,0.45) {$s=0$};
  \node[anchor=north] at (1.7,-0.08) {$(\textcolor{slab}{s},\textcolor{blue}{\theta},\textcolor{etac}{\eta})\in[0,L_j]\times[0,2\pi)\times[0,1]$};
\end{tikzpicture}}
  \label{fig:axis-angular}
\end{wrapfigure}

The field $R_j$ encodes the cross-sectional shape of the wrapper. Since $\GG_j$ lies strictly inside $\mathcal{W}_j$ by construction, $R_j$ is strictly positive over its entire domain.

We next define an axis-angular coordinate system using the guide as the cross-sectional centerline. Every point $\xx \in \mathcal{W}_j$ is represented by axis-angular coordinates $(s, \theta, \eta) \in [0, L_j] \times [0, 2\pi) \times [0, 1]$, defined by
\begin{equation}\label{eq:wd_axang}
    \xx = \GG_j(s) + \eta R_j(s, \theta)\ee_j(s, \theta),
\end{equation}
where $s$ is the arc length of the closest point on $\GG_j$, $\theta$ is the angular coordinate in $\mathcal{P}_j(s)$, and $\eta \in [0, 1]$ is the normalized radial coordinate, scaled so that $\eta = 1$ lies on $\partial\mathcal{W}_j$.

\paragraph{Wrapper deformation.}
Each scalp root $\rr_i \in \mathcal{R}_j$ is assigned canonical coordinates $(0, \theta_i, \eta_i)$ at the base of guide $j$. We then parallel-transport the canonical pair $(\theta_i, \eta_i)$ unchanged along $\GG_j$. Since Eq.~\eqref{eq:wd_axang} provides a continuous mapping from axis-angular coordinates to points inside $\mathcal{W}_j$, we evaluate this mapping at every sample along the strand to obtain
\begin{equation}\label{eq:wd_strands}
     \SS_i(s) = \GG_j(s) + \eta_i R_j\bigl(s, \theta_i\bigr)\ee_j\bigl(s, \theta_i\bigr).
\end{equation}

\begin{figure}
    \centering
    \includegraphics[width=0.9\linewidth]{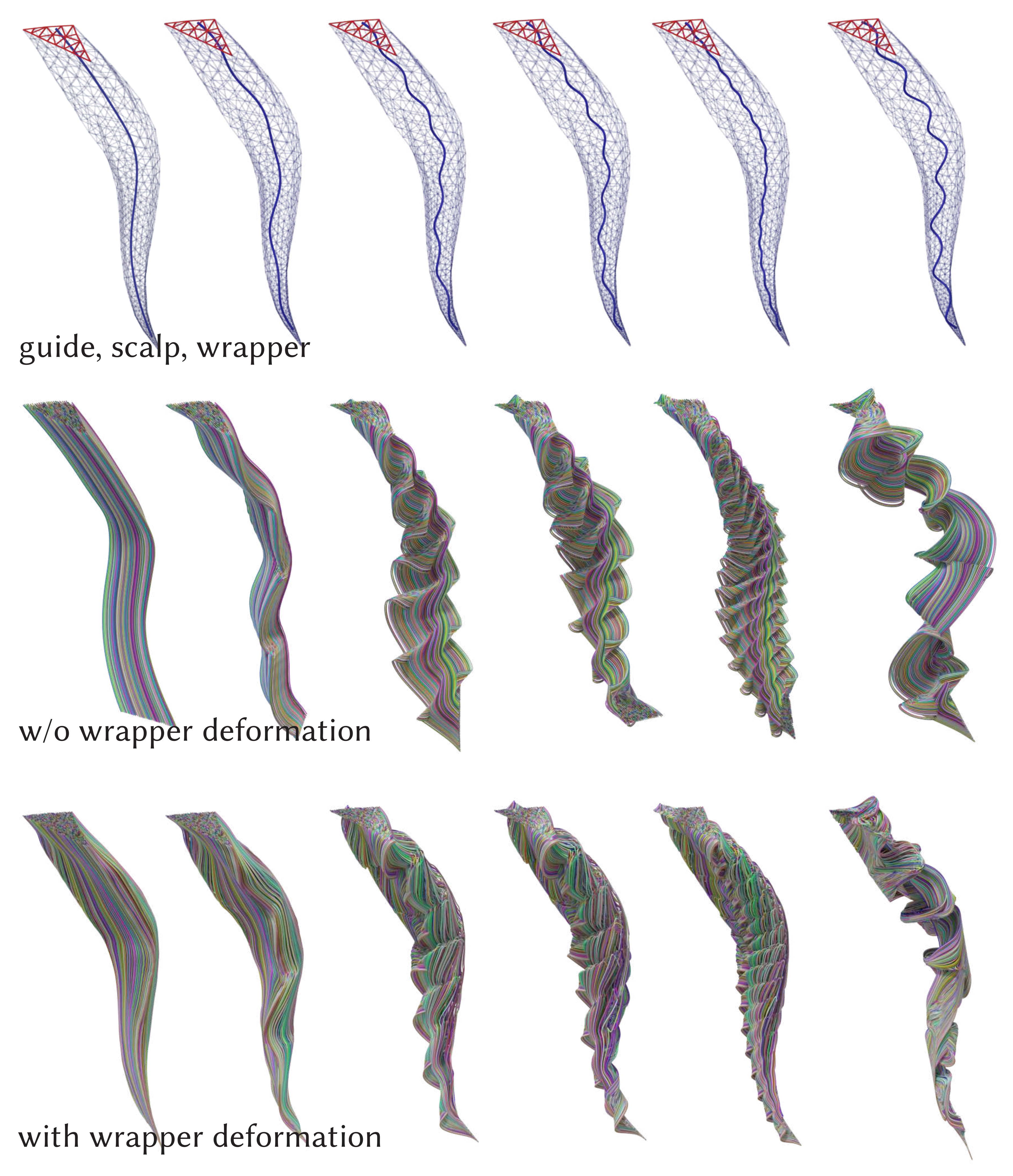}
    \caption{\textbf{Wrapper deformation.} The figure illustrates our wrapper-deformation experiments. The first row shows the scalp (red wireframe), guide strands (blue), and wrapper (blue wireframe). The next two rows show interpolation results using different frame-transport methods, without (row $2$) and with (row $3$) wrapper deformation.}
    \Description{}
    \label{fig:wd_exps}
\end{figure}

This construction generates scalp-connected strands that remain inside the wrapper while conforming to its local cross-sectional geometry. \autoref{fig:wd_exps} shows that the proposed wrapper deformation is robust across challenging guide configurations and produces plausible results that preserve the characteristic shape of the wrapper.

\begin{table*}[t]
  \centering
  \caption{\textbf{Hyperparameters.} The single parameter set used for every result in this paper. The head mesh is normalized to unit scale, so lengths are in head units.}
  \label{tab:hparams}
  \footnotesize
  \setlength{\tabcolsep}{5pt}
  \begin{tabular}{@{}llp{3.4cm}p{6.9cm}@{}}
    \toprule
    Stage & Symbol & Value & Comment \\
    \midrule
    Guide extraction (\autoref{sec:extraction}) & $|\SS_i|$ & 32, 64 & samples per guide and per dense strand for base hair and for groomed hair; 128 for the strongly curved in-the-wild cards of \autoref{sec:generalization} \\
     & guides per card & 1 or 3 & one centerline for thin cards, three for wide cards \\
    \midrule
    Guide binding (\autoref{sec:binding}) & $N^{\text{root}}$ & 30K & Poisson root candidates on the scalp \\
     & $w^{\text{dist}}, w^{\text{ang}}$ & 1, 10 & distance and normal-alignment weights in Eq.~\eqref{eq:dgeo} \\
     & $\tau$ & $\cos(3\pi/4)$ & opposing-tangent threshold for parting detection \\
    \midrule
    Root tracing (\autoref{sec:tracing}) & $N^{\text{extra}}$ & 200 & FPS budget of extra roots \\
     & $\epsilon_\text{offset}$ & $2\times10^{-3}$ & maximum layered-offset depth (\autoref{fig:layer_offset}) \\
     & $K$ & $0.2\,|\SS_i|$ & root-region samples smoothed by Eq.~\eqref{eq:smooth} \\
     & $w^{\text{lap}}, w^{\text{ori}}, w^{\text{proj}}, w^{\text{len}}$ & 10, 10, 1, 1 & smoothing weights in Eq.~\eqref{eq:smooth}; 20 iterations with step 0.1 \\
    \midrule
    Wrapper relaxation (\autoref{sec:relaxation}) & $\epsilon^{\text{thickness}}$ & $2\times10^{-3}$ & card slab thickness of $\Omega$ \\
     & $h_\Omega$ & $128^3$ & voxel grid of $\Phi_\Omega$, $\Phi_\mathcal{B}$, $\rho$, and $\OO$ \\
     & $K_\text{relax}, \Delta r$ & 30, $10^{-3}$ & expansion iterations and radial step (\autoref{alg:tube_relaxation}); spindles start at 10\% of the target ring \\
     & $\lambda_{\text{cap}}$ & 1.5 & a ring vertex never moves farther from the centerline than $\lambda_{\text{cap}}$ times its target-ring radius \\
     & $\xi$ & $3\times10^{-4}$ & ACCD safety margin \\
     & $K_\text{smooth}, \alpha$ & 30, 0.5 & longitudinal Laplacian sweeps and blend weight (\autoref{alg:tube_relaxation}) \\
     & $M/h^2, w^{\text{len}}, w^{\text{ang}}$ & $10^3$, $10^3$, $10^2$ & XPBD stiffness of the momentum, length, and angle terms (\autoref{apdx:relaxation}); 50 Gauss--Seidel sweeps \\
    \midrule
    Full hair synthesis (\autoref{sec:full-hair}) & root density & $10^6$ per unit area & scalp sampling density of the final strands \\
     & $\sigma_t, \sigma_b, \sigma_n$ & 2, 0.5, 0.5 voxels & Gaussian footprint of the orientation field $\OO$ and the target density $\rho^\star_{\OO}$ \\
     & $w^{\text{shape}}, w^{\text{den}}, w^{\text{wrap}}, w^{\text{bust}}$ & 10, 1, $10^4$, $10^4$ & relaxation weights in Eq.~\eqref{eq:relaxation_terms}; $w_j$ ramps up over the first joints \\
     & $\epsilon$ & $2\times10^{-3}$, 0 & wrapper and bust margins in Eq.~\eqref{eq:relaxation_terms} \\
     & Adam & $\alpha = 5\times10^{-3}$, $\beta = (0.9, 0.999)$ & run until the loss drops to 10\% of its initial value \\
    \bottomrule
  \end{tabular}
\end{table*}

\section{Helix Operator with Rotational Transport}\label{apdx:helix_operator}

As shown in \autoref{fig:wd_exps}, instead of synthesizing curly hair by applying a post-hoc operator to straight guides, we find that it is more effective to modify the initialization stage of full-hair synthesis by directly deforming the guides inside the wrapper to produce curly or wavy structure. Initialization then proceeds by applying rotational transport to generate the strand bundle corresponding to the second row of \autoref{fig:wd_exps}, after which each strand is further deformed to conform to the wrapper mesh. All subsequent stages, including orientation-field construction, density-field construction, and relaxation, remain unchanged. This design ensures that the synthesized strands preserve the intended hairstyle shape.

We expose three scalar parameters per guide. Let $\xi = s / L_j \in [0, 1]$ denote the normalized arc length along guide $j$, where $\xi = 0$ corresponds to the scalp root and $\xi = 1$ to the tip. The turn count $\tau$ specifies the number of complete $2\pi$ revolutions accumulated from $\xi = 0$ to $\xi = 1$. The ramp exponent $\beta > 0$ controls how this rotation is distributed along the guide. The radius $r$ gives the lateral amplitude of the helix in world units. The angular displacement at $\xi$ is
\begin{equation}\label{eq:helix_angle}
    \phi(\xi) \;=\; 2\pi\,\tau\,\xi^{\beta},
\end{equation}
and the deformed guide is defined as
\begin{equation}\label{eq:helix_guide}
    \tilde{\GG}_j(\xi) = \GG_j(\xi) + r\bigl(\cos\phi(\xi)\,\NN_j(\xi) + \sin\phi(\xi)\,\BB_j(\xi)\bigr),
\end{equation}
where $\{\NN_j(\xi), \BB_j(\xi)\}$ is the parallel-transport cross-sectional frame anchored at the root.

\paragraph{Rotational transport.}
Once the helical guide $\tilde{\GG}_j$ is defined, we transport each strand's cross-sectional offset along the deformed tangent field
$\tilde{\TT}_j(\xi):= \tilde{\GG}_j'(\xi) / \|\tilde{\GG}_j'(\xi)\|$
using incremental Rodrigues rotations.

We sample the guide at $0 = \xi_0 < \xi_1 < \dots < \xi_K = 1$, and let $\bm{\delta}_i^{(0)}$ denote the root offset of strand $i$, i.e., the in-plane component of $\bm{r}_i - \GG_j(0)$ in the rest frame at $\xi_0$. For each segment $k$, we construct the rotation that aligns $\tilde{\TT}_j(\xi_k)$ with $\tilde{\TT}_j(\xi_{k+1})$:
\begin{equation}\label{eq:helix_rod}
\begin{aligned}
    \hat{\mathbf{\omega}}_k &= \frac{\tilde{\TT}_j(\xi_k) \times \tilde{\TT}_j(\xi_{k+1})}{\|\tilde{\TT}_j(\xi_k) \times \tilde{\TT}_j(\xi_{k+1})\|},
    \\
    \alpha_k &= \arccos\bigl(\tilde{\TT}_j(\xi_k) \cdot \tilde{\TT}_j(\xi_{k+1})\bigr),
    \\
    \hat{\RR}_k &= \mathrm{Rodrigues}(\hat{\mathbf{\omega}}_k, \alpha_k),
\end{aligned}
\end{equation}
and propagate the offset via
\begin{equation}\label{eq:helix_offset}
    \bm{\delta}_i^{(k+1)} = \hat{\RR}_k\,\bm{\delta}_i^{(k)}.
\end{equation}

This procedure performs rotation-minimizing parallel transport along $\tilde{\TT}_j$, lifted from the centerline to the cross-section. As a result, each strand preserves a constant signed angle relative to its initial frame while the frame rotates with the tangent. The interpolated strand position at sample $k$ is then
\begin{equation}\label{eq:helix_strand}
    \xx_i^{\text{interp}}(\xi_k) = \tilde{\GG}_j(\xi_k) + \bm{\delta}_i^{(k)}.
\end{equation}

This yields the strand bundle filling the tube around the deformed guide, corresponding to the second row of \autoref{fig:wd_exps}. We then feed $\xx_i^{\text{interp}}$ into the per-cross-section deformation described in \autoref{apdx:wrapper_deformation}, specifically Eq.~\eqref{eq:wd_strands}, to warp the bundle so that it matches the cross-sectional shape of $\mathcal{W}_j$. Together, Eq.~\eqref{eq:helix_rod}--\eqref{eq:helix_offset} place the helix-mode strands inside the deformed tube, while Eq.~\eqref{eq:wd_strands} conforms them to the wrapper geometry.

\section{Interpolation Settings}\label{apdx:houdini}
The prism-based and clump-based baselines of \autoref{fig:interp} are generated with the Hair Generate node of Houdini from the same guide set as ours, with the settings of \autoref{tab:houdini}, which we tuned by hand for each baseline to obtain its best result; the generated strands are not edited afterwards. The two differ only in the four rows below the rule: the prism-based variant blends up to three guides linearly, whereas the clump-based variant follows a single guide by extrusion.
\begin{table}[ht!]
  \centering
  \caption{\textbf{Interpolation settings.} Parameters of the Houdini Hair Generate node for the two baselines of \autoref{fig:interp}. The rows above the rule are shared; the rows below it differ.}
  \label{tab:houdini}
  \footnotesize
  \begin{tabular}{@{}lll@{}}
    \toprule
    Parameter & Prism-based & Clump-based \\
    \midrule
    Hair count           & 99,000 & 99,000 \\
    Scatter seed         & 5.35 & 5.35 \\
    Relax iterations     & 20 & 20 \\
    Skin guide mode      & Match by Guide Id & Match by Guide Id \\
    Influence decay      & 1 & 1 \\
    Maximum guide angle  & $90^\circ$ & $90^\circ$ \\
    Unguided hairs       & off & off \\
    \midrule
    Blending method      & Linear Blend & Extrude And Blend \\
    Maximum guide count  & 3   & 1 \\
    Influence radius     & 0.1 & 0.3 \\
    Clump crossover      & 0.3 & 0.1 \\
    \bottomrule
  \end{tabular}
\end{table}

\section{Hyperparameters}\label{apdx:hparams}
All results use the parameter set in \autoref{tab:hparams}.
\section{50K Strand-based Hair Dataset}\label{apdx:dataset}
The dataset is available at \url{https://huggingface.co/datasets/HairCS2027/HairCS}. In the following, we provide more details about the dataset.

\begin{table}[t]
  \centering
  \caption{\textbf{Dataset composition.} Hairstyles per label class, split by how they were produced: reconstructed from cards (base) or blended (\autoref{sec:blending}), each without (straight) or with the curly/helix operator (helix).}
  \label{tab:dataset}
  \footnotesize
  \setlength{\tabcolsep}{5pt}
  \begin{tabular}{@{}lrrrrr@{}}
    \toprule
    & \multicolumn{2}{c}{Base} & \multicolumn{2}{c}{Blend} & \\
    \cmidrule(lr){2-3}\cmidrule(lr){4-5}
    Class & straight & helix & straight & helix & Total \\
    \midrule
    short    &   570 &   746 &  8,585 &  8,585 & 18,486 \\
    bob      &   250 &   332 &     66 &     66 &    714 \\
    shoulder &   170 &   250 &  7,559 &  7,559 & 15,538 \\
    long     &   632 &   601 &  8,226 &  8,226 & 17,685 \\
    gather   &   530 &   383 &      0 &      0 &    913 \\
    \midrule
    Total    & 2,152 & 2,312 & 24,436 & 24,436 & 53,336 \\
    \bottomrule
  \end{tabular}
\end{table}

All hairstyles are grown on the same bust $\mathcal{B}$ and scalp $\mathcal{S}$, which ship with the dataset as OBJ meshes, so every hairstyle lives in one coordinate frame and any two of them can be combined without alignment. Each hairstyle is stored as a single array of $60{,}000 \times 64 \times 3$ half-precision (\texttt{float16}) coordinates: 60K strands with roots scattered uniformly over the scalp region, each resampled to 64 points ordered from root to tip. We chose half precision because its rounding error stays below $0.5$\,mm, about $0.2\%$ of the head height and far below the strand spacing. A hairstyle takes about 18\,MB as a\lastpagecolumnbreak{} compressed \texttt{npz} file, and the complete set of 53,336 hairstyles about 0.93\,TB.

The set comes in 13 versions that differ in how the hair was produced (\autoref{tab:dataset}). Of these, 4,464 hairstyles are direct outputs of our pipeline, either the reconstructed base hair or that hair enriched with grooming operators: the curly/helix operator of \autoref{apdx:helix_operator} in two parameter profiles, and a scale-and-fuzz modifier (\autoref{fig:groom}). For the helix operator, the turn count and the radius of each guide follow its arc length: a guide of median length receives the base values $\tau$ and $r$, the shortest guide $\tau - \Delta\tau$ and $r - \Delta r$, the longest $\tau + \Delta\tau$ and $r + \Delta r$, with linear interpolation in length on each side of the median. The first profile uses $\tau = 10 \pm 6$, $r = 0.004 \pm 0.002$, and $\beta = 1.7$; the second uses $\tau = 16 \pm 10$, $r = 0.004 \pm 0.002$, and $\beta = 1.2$, with $r$ in head units. The remaining 48,872 hairstyles come from the hairstyle blending of \autoref{sec:blending} towards short, bob, shoulder, and long targets, and we again provide each of them both in its original form and with the helix operator applied. Every hairstyle carries one of five length/type labels, short, bob, shoulder, long, and gather (buns, knots, ponytails, and braids), so the set can be filtered by grooming operator, hair length, and hair type.

Every hairstyle comes with a rendered preview image. The repository also includes a lightweight strand viewer and a strand-to-surface tool that wraps a hairstyle into a watertight mesh, by voxel rasterization, a signed distance field, marching cubes, and isotropic remeshing, for downstream use.

Part of the dataset is shown in \autoref{fig:evaluation} and \autoref{fig:gallary}: the former holds the reconstructed base hairstyles of our evaluation, the latter a random sample of 300 hairstyles from the full set.

\end{document}